\documentclass[noeprint,twocolumn,aps,prb,showpacs,groupedaddress,superscriptaddress,nofootinbib,longbibliography]{revtex4-2} 
\usepackage{graphicx}  
\usepackage[dvipsnames]{xcolor} 
\usepackage{color}     
\usepackage{amsmath}   
\usepackage{amsfonts}  
\usepackage{bbold}
\usepackage{float}     
\usepackage[pdftex,linkcolor=blue,citecolor=blue,urlcolor=blue,colorlinks]{hyperref} 
\usepackage{outlines} 
\usepackage{lipsum}
\usepackage{bm}    
\usepackage{mathrsfs}
\usepackage{bigstrut}
\usepackage{enumerate}
\usepackage{etoolbox}
\usepackage[normalem]{ulem}

\newcommand{\appropto}{\mathrel{\vcenter{
  \offinterlineskip\halign{\hfil$##$\cr
    \propto\cr\noalign{\kern2pt}\sim\cr\noalign{\kern-2pt}}}}} 

\begin{document}
	
\title{
Fragmentation, Zero Modes, and Collective Bound States in Constrained Models
}

\author{Eloi Nicolau}
\affiliation{Institute of Science and Technology Austria, Am Campus 1, Klosterneuburg, 3400, Austria}
\author{Marko Ljubotina}
\affiliation{Technical University of Munich, TUM School of Natural Sciences, Physics Department, 85748 Garching, Germany}
\affiliation{Munich Center for Quantum Science and Technology (MCQST), Schellingstr. 4, München 80799, Germany}
\affiliation{Institute of Science and Technology Austria, Am Campus 1, Klosterneuburg, 3400, Austria}

\author{Maksym Serbyn}
\affiliation{Institute of Science and Technology Austria, Am Campus 1, Klosterneuburg, 3400, Austria}

\begin{abstract}
Kinetically constrained models were originally introduced to capture slow relaxation in glassy systems, where dynamics are hindered by local constraints instead of energy barriers. Their quantum counterparts have recently drawn attention for exhibiting highly degenerate eigenstates at zero energy -- known as \emph{zero modes} -- stemming from chiral symmetry. Yet, the structure and implications of these zero modes remain poorly understood. In this work, we focus on the properties of the zero mode subspace in quantum kinetically constrained models with a $U(1)$ particle-conservation symmetry. We use the $U(1)$ East, which lacks inversion symmetry, and the inversion-symmetric $U(1)$ East-West models to illustrate our two main results. First, we observe that the simultaneous presence of constraints and chiral symmetry generally leads to a \emph{parametric increase} in the number of zero modes due to the fragmentation of the many-body Hilbert space into disconnected sectors. Second, we generalize the concept of compact localized states from single particle physics and introduce the notion of \emph{collective bound states}, a special kind of non-ergodic eigenstates that are robust to enlarging the system size. We formulate sufficient criteria for their existence, arguing that the degenerate zero mode subspace plays a central role, and demonstrate bound states in both example models and in a two-dimensional model, the $U(1)$ North-East, and in the pair-flip model, a system without particle conservation. Our results motivate a systematic study of bound states and their relation to ergodicity breaking, transport, and other properties of quantum kinetically constrained models. 
\end{abstract}
\maketitle

\section{Introduction}

Out-of-equilibrium properties of quantum many-body models have been one of the central questions in quantum physics in recent years. In this context, many such features can be linked to the spectral properties of the studied models, which are often strongly influenced by their symmetry structure~\cite{RMTWigner, RMTMehta} (e.g., in random matrix theory). The most commonly studied cases involve the presence or absence of time-reversal invariance. However, there also exist special symmetry classes that can lead to unique spectral features, such as protected states at zero energy, with Majorana zero modes being among the most notable examples~\cite{read2000,kitaev2001}.

Although Majorana zero modes~\cite{read2000,kitaev2001}, which require particle-hole symmetry, were mostly studied in the context of non-interacting systems~\cite{Sarma_2015, Tanaka_2024}, several early works demonstrated their stability to certain interactions~\cite{PhysRevB.86.115122,PhysRevB.99.195123,PhysRevD.101.066017}. More recently, exponentially degenerate  subspaces at zero energy were observed in the context of 
kinetically constrained and other spin and bosonic 
models~\cite{Turner2017,schecter2018,karle2021,gerken2023,theel2024}. 
These \emph{zero mode} (ZM) subspaces instead rely on the presence of chiral symmetry, which also imposes a symmetric energy spectrum, and can thus be thought of as many-body counterparts of a special symmetry class.

The presence of a subspace with a large degeneracy pinned to zero energy also warrants exploration from the perspective of the eigenstate thermalization hypothesis (ETH). The ETH conjectures that the expectation values of local observables over eigenstates coincide with the thermal expectation values at a corresponding temperature and puts further assumptions on the structure of the off-diagonal matrix elements of local operators between eigenstates~\cite{Deutsch1991,Srednicki1994,Rigol2008}. Thus, eigenstates from the ZM subspace, where the density of states typically peaks, should have observables that agree with thermal expectation values at infinite temperature. Although degeneracy prevents the immediate application of the ETH to the ZM subspace, it has been suggested that the entire subspace can feature deviations from thermal behavior~\cite{schecter2018}. Additional evidence for non-thermal behavior was given by the existence of exact~\cite{lin2018,lin2020,banerjee2021,surace2021,ivanov2025} or approximate~\cite{karle2021} area-law entangled eigenstates within the zero-mode subspace. 

Despite the progress mentioned above, the majority of the studied models feature a single conserved charge, the energy. However, many physically relevant models conserve additional charges, such as the total particle number, leading to additional symmetries. Such particle or spin-conserving models can also exhibit degenerate ZM subspaces. However, the effects of the $U(1)$ symmetry in such cases are not yet understood \cite{aditya2024}. In this work, we focus on the properties of $U(1)$ symmetric models with ZM subspaces and kinetic constraints. In particular, we present two main results: the parametric enhancement of the ZM subspace degeneracy and the existence of \emph{collective bound states} within the ZM subspace, both attributed to the non-trivial interplay of constraints and particle number conservation.

In order to illustrate our general results, we focus on two particular families of models dubbed the $U(1)$ conserving East and East-West models, which both have exponentially degenerate ZM subspaces due to chiral symmetry. Notably, the $U(1)$ East model breaks inversion symmetry, whereas the $U(1)$ East-West model (a generalization of the Gonçalves-Landim-Toninelli model~\cite{goncalves2009,Vasseur2021}) is inversion symmetric. First, we derive a lower bound for the size of the degenerate subspace and illustrate the mechanism of its enlargement in the presence of Hilbert space fragmentation~\cite{Moudgalya2022Review}. Then, we define the notion of collective bound states and demonstrate their existence in both models, highlighting the critical role played by the degenerate ZM subspace. These eigenstates are a special kind of Fock space or many-body cages~\cite{tan25,benami25,jonay2025}. Many-body cages are eigenstates that are localized in Fock space: they have a low number of product states contributing to the state and amplitude zero on the rest. Collective bound states are many-body cages that are not only localized in Fock space but also robust to increasing the size of the system. Finally, we discuss that combining bound states for smaller system sizes together leads to factorizable eigenstates -- featuring zero entanglement across certain cuts -- in larger systems.

For the inversion-breaking $U(1)$ East model, we demonstrate the intuitive principle behind the construction of such bound states and show that their number grows with system size. Moreover, we show that the inversion-breaking nature of the $U(1)$ East model enables bound states to ``poison'' the spectrum of the model with non-thermal factorizable eigenstates away from zero energy. For the case of the inversion symmetric East-West model, we present a specific MPO construction for collective bound states and discuss the generation of factorizable eigenstates, which in this case only arise within the ZM subspace. In addition, we show that collective bound states are not just a special feature of the $U(1)$ East and East-West models by discussing two more models: the North-East model, a two-dimensional system, and the pair-flip model, therefore showing that collective bound states can also arise in two dimensions and in systems without particle conservation.

The remainder of this paper is organized as follows. In Sec.~\ref{SecChiralModels}, we review chiral symmetry in models without $U(1)$ symmetry and then introduce the $U(1)$ conserving East and East-West models. This is followed by Sec.~\ref{SecCountingZM}, where we lower bound the number of ZMs for a generic $U(1)$ conserving model with chiral symmetry and compare with the numerical results for the $U(1)$ East and East-West models. In Sec.~\ref{SecBoundStatesAndFactorizableStates}, we present our results on collective bound states and factorizable ZMs. Specifically, in Sec.~\ref{SecDefinitions}, we present a generic definition, construction, and sufficient conditions for the existence of these states. Then, in Sec.~\ref{SecEast} and Sec.~\ref{SecEastWest}, we perform these constructions for the $U(1)$ East and East-West models, respectively. In Sec.~\ref{SecGeneralization}, we discuss the North-East and pair-flip models that generalize our results to two-dimensions and to models without particle conservation. Finally, in Sec.~\ref{SecDiscussion}, we discuss the implications of our results, their potential extensions, and relevant open questions. 

\section{Particle conserving chiral models with kinetic constraints}\label{SecChiralModels}

In this section, we introduce two different classes of one-dimensional particle-conserving interacting models that feature a ZM subspace. First, we review how these degenerate subspaces can arise in spin models without $U(1)$ symmetry due to the presence of chiral symmetry. Then, we introduce two specific models with both chiral and $U(1)$ symmetries that will be used to illustrate the general results obtained in our work. 

\subsection{Chiral symmetry without $U(1)$ conservation}
Zero mode (ZM) subspaces can arise due to the presence of chiral symmetry, 
defined via a unitary and Hermitian operator $\hat{\mathcal{C}}$ that anti-commutes with the Hamiltonian, $\{\hat{H},\hat{\mathcal{C}}\}=0$. Due to this anti-commutation relation, chiral symmetry does not lead to an additional conserved quantity, but instead, it leads to the symmetry of the energy spectrum around $E=0$. Specifically, for every eigenstate at energy $E$, $|\psi_E\rangle$, we can obtain an eigenstate at energy $-E$ via the action of $\hat{\mathcal{C}}$ operator, $|\psi_{-E}\rangle = \hat{\mathcal{C}}|\psi_E\rangle$. 

Let us first review the simplest class of interacting models with chiral symmetry by analyzing a spin-1/2 chain. Consider the family of interacting Hamiltonians constructed by summing strings of Pauli operators that contain an odd number of $\hat{\sigma}^x$ and an arbitrary number of $\hat{\sigma}^z$ matrices acting on different sites. The Hamiltonian of this large class of models reads,
\begin{equation}
    \hat{H} = \sum_i (a\hat{\sigma}^x_i+b\hat{\sigma}^z_{i-1}\hat{\sigma}^x_i+c\hat{\sigma}^x_{i-1}\hat{\sigma}^x_i\hat{\sigma}^x_{i+1}+\ldots),
\end{equation}
where  $\{a,b,c,\ldots\}$ are arbitrary coefficients and we denote the set of Pauli matrices acting on a local spin $i$ by $\hat{\sigma}^{x,y,z}_i$. In this case, the operator $\hat{\mathcal{C}}$ corresponds to the total spin parity along $z$-direction, $\hat{\mathcal{C}}=\prod_i \hat{\sigma}^z_i$~\cite{karle2021}. Although these models are not constrained, there are kinetically constrained models, such as the PXP model~\cite{Turner2017,schecter2018}, that also anti-commute with this chiral operator $\hat{\mathcal{C}}$.

In such a class of models, chiral symmetry results in an intuitive splitting of all product states in the Hilbert space into two classes: those with even parity and those with odd parity with respect to the total number of $\downarrow$-spins. As all terms in the Hamiltonian perform an odd number of spin flips, the Hamiltonian only connects product states with opposite parities, and can thus be written in a block off-diagonal form
\begin{equation}\label{Eq:chiral}
    \hat{H} = \begin{pmatrix}
        0 &C \\
        C^\dagger & 0
    \end{pmatrix},
\end{equation}
by adequately ordering the product states. The structure of the Hamiltonian matrix induced by chiral symmetry exposes the origin of the ZM subspace. The difference between the numbers of rows and columns in $C$ corresponds to the difference in the numbers of states in odd and even-parity sectors, $N_o$ and $N_e$. This mismatch puts a lower bound on the dimension of the kernel of $\hat{H}$, i.e., on the number of ZMs,
\begin{equation} \label{Eq:dim-ker}
    \mathop{\rm dim} \mathop{\rm ker} \hat{H} \geq {\cal M} = |N_e-N_o|,
\end{equation}
because it limits the maximum number of linearly independent rows in $\hat{H}$. This is the same mechanism that leads to non-interacting localized states on defects in bipartite lattices~\cite{PhysRevB.34.5208,PhysRevB.49.3190}. Note that ZMs here simply refer to eigenstates with energy zero. They are different from strong zero modes, operators that commute with a Hamiltonian in the thermodynamic limit and do not commute with a discrete symmetry of the system, making the energy spectrum composed of pairs of nearly degenerate eigenstates \cite{fendley2016}.

\subsection{Chiral symmetry with $U(1)$ conservation}
The models introduced above, where chiral symmetry relies on total spin parity, do not have $U(1)$ symmetry. In order to introduce a $U(1)$ symmetric model, we draw inspiration from noninteracting systems. The simplest chiral model is given by a single particle on a one-dimensional lattice with nearest neighbor hopping, $\hat{H} = \sum_i (\hat{c}^\dagger_{i+1} \hat{c}_i+\text{h.c.})$. We assume particles to be hard-core bosons, such that there are only two states per site\footnote{Note that a such a hard core boson model can be mapped to a free fermion model through a Jordan-Wigner transformation.}. Here, the chiral operator that anti-commutes with the Hamiltonian may be constructed as a particle number parity of one \emph{sublattice}, either the set of even or odd sites. For the even sites, the operator reads
\begin{equation} \label{eq:calC}
\hat{\mathcal{C}}=\prod_{i=1}^{\lfloor L/2\rfloor}(2\hat{n}_{2i}-1),
\end{equation}
where $\hat{n}_i = \hat{c}^\dagger_i \hat{c}_i$ is the particle occupation number on each site $i$, which yields values 0 or 1 for hard-core bosons and $L$ is the number of sites. The noninteracting Hamiltonian only couples states with opposite parities: those with an odd number of particles in one sublattice to those with an even number of particles. Adding generic interactions to the above Hamiltonian violates the anti-commutation relation $\{\hat{H},\hat{\mathcal{C}}\}=0$ that defines the chiral symmetry. Indeed, any diagonal density term commutes with $\hat{\mathcal{C}}$, and thus yields a nonzero anti-commutator, $\{\hat{H},\hat{\mathcal{C}}\}\neq0$. However, chiral symmetry is robust to the addition of correlated hopping terms such as $\hat{n}_{i+2}(\hat{c}^\dagger_{i+1} \hat{c}_i+\text{h.c.})$, as those are off-diagonal and only couple product states with opposite parities. 

The arguments above show that hard-core bosons with correlated nearest-neighbor hopping provide a family of models with chiral symmetry given by the operator~\eqref{eq:calC}. One might simplify these models by ignoring the uncorrelated hopping terms, which leads to a family of kinetically constrained models. In particular, we will focus on the particle conserving East model and the particle conserving East-West model, denoted as $U(1)$ East and $U(1)$ East-West, which describe hard-core bosons in a one-dimensional lattice. These models consist of a nearest neighbor hopping of hard-core bosons, that is enabled by the presence of certain particle configurations on $r$ nearby sites and have chiral symmetry generated by the operator~\eqref{eq:calC}. Both models were studied in the literature, however focusing on different aspects: the $U(1)$ East model was first proposed in the context of localization physics~\cite{Brighi20}, and Hilbert space fragmentation~\cite{brighi2023,ganguli2025,aditya2025}, and both the $U(1)$ East and East-West models were studied in the context of particle number transport~\cite{brighi2023,brighi2024}.

\begin{figure}[t]
\includegraphics[width=\columnwidth]{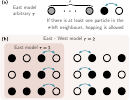}
	\caption{Illustration of the kinetic constraints of the $U(1)$ East and East-West models. Allowed hoppings for a kinetic constraint (a) of arbitrary range in the $U(1)$ East model and (b) of range $r=2$ in both models.}
	\label{FigModel}
\end{figure}

\begin{figure*}[t]
\includegraphics[width=2\columnwidth]{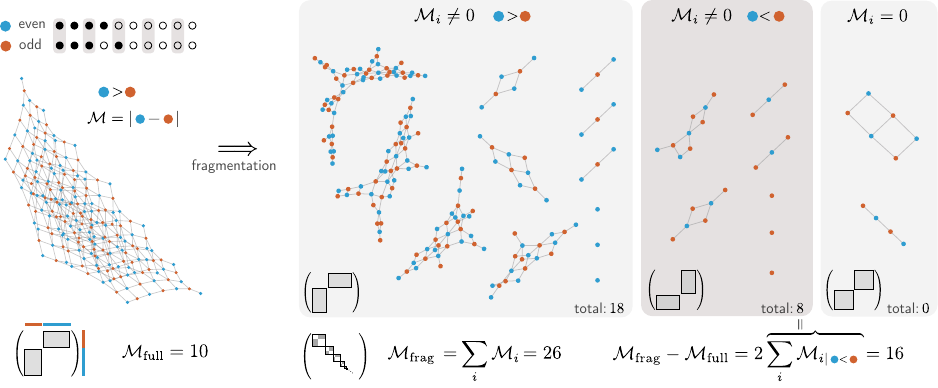}
    \caption{Adjacency graphs for the $U(1)$ East model with $N=4$ particles on $L=10$ sites before [Eq.~\eqref{EqHamiltonianEast} without the constraint operator $\hat{\mathcal{K}}_{i,r}$] and after fragmentation. The mismatch $\mathcal{M}$ between the number of blue vertices (with an even number of particles in the odd sublattice) and orange vertices (with an odd number) is indicated below. Fragmentation enhances the mismatch considerably, and thus it increases the lower bound on the number of ZM. The increase is determined by those sectors where the dominant set of vertices is the opposite of the one before fragmentation.}
	\label{FigFragmentation}
\end{figure*}

Specifically, we define the \emph{range-$r$ East model}~\cite{brighi2023} by the following Hamiltonian, 
\begin{equation}\label{EqHamiltonianEast}
	\begin{aligned}
		\hat{H}_{r,E}&=\sum_{i=r+1}^{L-1} \hat{\mathcal{K}}_{i, r}\left(\hat{c}_{i+1}^{\dagger} \hat{c}_i+\text {h.c.}\right),\\
		\hat{\mathcal{K}}_{i, r}&=\sum_{\ell=1}^r t_{\ell} \hat{n}_{i-\ell} \prod_{j=i-\ell+1}^{i-1}\left(1-\hat{n}_j\right),
	\end{aligned}
\end{equation}
where the projector $\hat{\mathcal{K}}_{i, r}$ enforces the constraint that a particle can hop if and only if there is at least one particle on the $r$ left neighboring sites, see Fig.~\ref{FigModel}(a). This projector has $r$ free parameters $t_\ell$ that specify the hopping amplitudes enabled by a neighboring particle at a certain distance. For all these models, where the hopping is enabled by the presence of particles only to the left of the moving particle, the inversion symmetry is strongly broken.
In the \emph{range-$r$ East-West model}~\cite{brighi2024}, inversion symmetry is restored by adding a spatially inverted Hamiltonian, which allows hopping conditioned on the presence of particles to the right, see Fig.~\ref{FigModel}(b). This leads to the total Hamiltonian 
\begin{equation}\label{eq:east-west}
    \hat{H}_{r,EW}=\hat{H}_{r,E}+\hat{H}_{r,W},
\end{equation}
where $\hat{H}_{r,W}$ is the mirror-reflection of $\hat{H}_{r,E}$. 

Although our results below are valid for general $r$, we will mostly focus on $r=2$ range models. At this range, setting the values of all hoppings $t_\ell=1$, we obtain the following expressions for the $U(1)$ East Hamiltonian and its mirror image,
\begin{eqnarray}\label{eq:r=2-east}
\hat{H}_{2,E}&\!=\!&\sum_{i=2}^{L-1}(\hat{n}_{i-1}\!+\!\hat{n}_{i-2}\!-\!\hat{n}_{i-1} \hat{n}_{i-2})(\hat{c}_i^{\dagger} \hat{c}_{i+1}\!+\!\text {h.c.}),\\
\label{eq:r=2-west}
\hat{H}_{2,W}&\!=\!&\sum_{i=1}^{L-2}(\hat{c}_i^{\dagger} \hat{c}_{i+1}\!+\!\text {h.c.})
\left(\hat{n}_{i+2}\!+\!\hat{n}_{i+3}\!-\!\hat{n}_{i+2} \hat{n}_{i+3}\right)\!,
\end{eqnarray}
where we fix $\hat{n}_{i<1}=\hat{n}_{i>L}=0$. The kinetic constraints in the $U(1)$ East model can freeze particles in a certain position through the lack of neighboring particles to their left. For example, the leftmost particle is always frozen for open boundary conditions, which we use throughout this work. In addition, the model exhibits Hilbert space fragmentation~\cite{Moudgalya2022Review}: the Hilbert space is composed of dynamically disconnected sectors of varied dimensions that increase exponentially in number with the size of the system. We fix a particle filling through $L=(r+1)N-r$ sites, where $N$ is the particle number, which naturally arises from this fragmentation structure (see Sec.~\ref{SecZMCountFragmentation}). In contrast, the $U(1)$ East-West model can only feature a few frozen states, which are dynamical sectors of dimension one, due to its less restrictive kinetic constraint. As we will discuss in the next section, fragmentation can have an important enhancement effect on the number of ZMs of the model. 

\section{Counting of zero modes}\label{SecCountingZM}
In this section, we derive a general lower bound for the number of ZMs in any $U(1)$ conserving model with chiral symmetry and two local degrees of freedom per site. Then, we discuss the enhancement effect of fragmentation on this lower bound and compare the numerical results for the $U(1)$ East and $U(1)$ East-West models to their analytical bounds. 

\subsection{General counting without fragmentation}
For a given number of particles $N$ and sites $L$, the Hamiltonian matrices of $\hat{H}_{E,r}$ and $\hat{H}_{EW,r}$ can be interpreted as adjacency matrices representing undirected graphs. Each product state corresponds to a vertex in the graph, while each matrix element corresponds to an edge. As a result of chiral symmetry, the matrices can be written in block off-diagonal form~(\ref{Eq:chiral}) by ordering the product states as $[\{|s_{e}^i\rangle\},\{|s_o^i\rangle\}]$ or $[\{|s_{o}^i\rangle\},\{|s_e^i\rangle\}]$, where $\{|s_e^i\rangle\}$ ($\{|s_o^i\rangle\}$) are the product states with an even (odd) number of particles on the even sites. Consequently, the graph is bipartite or two-colorable, with two sets of vertices $\{|s_{e}^i\rangle\}$ and $\{|s_o^i\rangle\}$ such that edges only exist between these two subsets of vertices. As discussed in the previous section,  any interaction term that would couple a product state to itself, yielding a diagonal term, would break chiral symmetry. 

The mismatch between the number of states $\{|s_e^i\rangle\}$ and $\{|s_o^i\rangle\}$, ${\cal M} = |N_e-N_o|$, will give a lower bound to the dimension of the kernel of the matrix, i.e., the number of ZMs (see Eq.~(\ref{Eq:dim-ker}) and illustration in Fig.~\ref{FigFragmentation}). Specifically, consider a bipartite lattice where the edges denote the hopping terms, with $L_A$ sites in one sublattice and $L_B$ sites in the other ($L=L_A+L_B$), and two local states per site. Given a $U(1)$ symmetry related to particle conservation, the Hilbert space dimension for each particle sector is given by the binomial coefficient $D = \binom{L}{N}=\frac{L!}{N!(L-N)!}$. Then, the mismatch between the number of odd and even product states in this sector reads
\begin{equation}\label{EqMismatch}
\mathcal{M}=\left|\sum_{n={\rm max}(0,N-L_B)}^{{\rm min} (N,L_A)}(-1)^n\begin{pmatrix}L_A\\n\end{pmatrix}\begin{pmatrix}L-L_A\\N-n\end{pmatrix}\right|,
\end{equation}
where the sum is taken over $n$, the number of particles in the $A$ sublattice. The lower bound on the number of ZMs ($\mathcal{M}$) grows exponentially with the number of particles, in a modulated manner, which is dependent on the particle filling. In Fig.~\ref{FigMismatch}, we represent $\mathcal{M}$ as a function of the number of particles for a 1D bipartite lattice with particle filling determined by the relation $L=3N-2$. We note that for such a fixed relation between system size and particle number, the lower bound on the number of ZMs simplifies to
\begin{equation}\label{EqMismatchFit}
    \mathcal{M}(N)=\left\{\begin{matrix}2\binom{{3N}/{2}-1}{N}&\quad \text{for}~ N~\text{even}\\3\binom{3N/{2}-5/2}{N-1}&\quad \text{for}~N>1~\text{odd}\end{matrix}\right.,
\end{equation}
which scales asymptotically as $\mathcal{M}(N)\appropto{2^{1-N}3^{3N/2}}/{\sqrt{6\pi N}}$ for even $N$, revealing exponential growth. Note that the relation between even and odd particle numbers is simply $\mathcal{M}_{\rm odd}(N)=\frac{3}{2}\mathcal{M}_{\rm even}(N-1)$. 
Let us now consider the effect of constraints on this bound. 

\subsection{Additional zero modes from fragmentation}\label{SecZMCountFragmentation}
For the $U(1)$ East model, the kinetic constraints are known to induce classical Hilbert space fragmentation. When a system is classically fragmented, the dynamically disconnected sectors of the Hilbert space are visible in the product state basis. Alternatively, if the sectors are only apparent in an entangled basis, the system is said to present quantum Hilbert space fragmentation \cite{Moudgalya2022}. In our case, the fragments appear through two mechanisms.
First, due to the choice of boundary condition, the leftmost particle is always frozen, as there are no particles to its left that can enable hopping. Therefore, product states where the leftmost particle occupies different positions are dynamically disconnected. Second, any particle preceded by $r$ empty sites and with too few particles to its left will also be frozen and thus, it will generate further fragmentation (e.g., for $r=1$, the third particle in this configuration is locked $|\bullet\bullet\circ\circ\circ\bullet\bullet\circ\rangle$). In contrast, the $U(1)$ East-West model is dynamically connected for $L<(r+2)N-(r+1)$. For more dilute particle fillings, only isolated frozen states appear, as two close particles can always propagate indefinitely (e.g., for $r=1$, $|\circ\bullet\bullet\circ\circ\rangle$ $\rightarrow$ $|\circ\bullet\circ\bullet\circ\rangle$ $\rightarrow$ $|\circ\circ\bullet\bullet\circ\rangle$). 

Adding a kinetic constraint removes matrix elements from the Hamiltonian, or, equivalently, in the graph description, this removes edges from the graph. If the system becomes classically fragmented, the Hamiltonian can be written in a block diagonal form in the product state basis, as the graph is no longer connected. Since the graph remains bipartite, i.e., chiral symmetry is preserved, each diagonal block retains the off-diagonal structure given by Eq.~(\ref{Eq:chiral}). Then, the lower bound to the number of ZMs for the entire Hilbert space comprises the contributions from all sectors $i$, $\mathcal{M}_{\rm frag}=\sum_{i}\mathcal{M}_{i}$, such that $\mathcal{M}\leq\mathcal{M}_{\rm frag}\leq D$, where $\mathcal{M}$ is the mismatch before fragmentation given by Eq.~(\ref{EqMismatch}). The increase in this lower bound is given by 
\begin{equation}\label{Eq:bound-flipped}
    \mathcal{M}_{\rm frag}-\mathcal{M}=2\sum_i\mathcal{M}_{i|\rm flipped},
\end{equation} 
where the sum includes all sectors where the largest set of vertices before fragmentation becomes the one with the fewest vertices in that sector. Figure~\ref{FigFragmentation} shows the colored adjacency graphs of the $U(1)$ East model for $N=4$ particles on $L=10$ sites before fragmentation [Eq.~\eqref{EqHamiltonianEast} without the constraint operator $\hat{\mathcal{K}}_{i,r}$] and after, along with the corresponding mismatch.  

\begin{figure}[t]
    \includegraphics[width=\columnwidth]{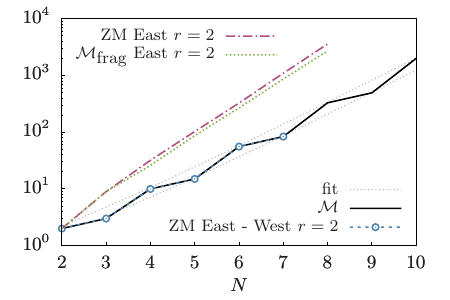}
    \caption{Number of ZMs in the $U(1)$ East and East-West models for $r=2$ and $L=3N-2$ as function of the number of particles $N$, and the analytical lower bound given by the mismatch $\mathcal{M}$, Eq.~\eqref{EqMismatch}, and the asymptotic expansion of Eq.~\eqref{EqMismatchFit}. Due to chiral symmetry, the number of ZMs increases exponentially, while the fragmentation in the $U(1)$ East model yields a faster exponential growth given by $\mathcal{M}_{\rm frag}$, which tightly bounds the number of ZMs from below.}
    \label{FigMismatch}
\end{figure}

The mismatch $\mathcal{M}$ or $\mathcal{M}_{\rm frag}$ provides a lower bound on the number of ZMs, which is generally not saturated, as nothing prevents the Hamiltonian matrix from having additional linearly dependent rows. One mechanism to produce additional ZMs is the presence of local reflection symmetries in the many-body graph (see Fig.~\ref{FigBoundstate} for $r=2$). Similar to the fragmentation mechanism, one must first resolve these symmetries, which yields a block diagonal structure for the Hamiltonian, before analyzing the mismatch in each block. Figure~\ref{FigMismatch} compares the number of ZMs in the $U(1)$ East and $U(1)$ East-West $r=2$ models for the particle filling determined by $L=3N-2$, against the mismatch before fragmentation predicted by Eq.~\eqref{EqMismatch} as a function of the number of particles. While the $U(1)$ East-West model saturates the $\mathcal{M}$ bound, the $U(1)$ East model presents a much faster exponential growth compared to the lower bound. Figure~\ref{FigMismatch} reveals that the bound that takes fragmentation into account, $\mathcal{M}_\text{frag}$, provides a much better approximation to the number of ZMs in $U(1)$ East model, however this model features extra ZMs even with respect to this bound. In the next section, we provide additional insights into the structure of the ZM subspace, demonstrating that many ZMs can be understood as originating from collective bound states. 

\section{Constructing factorizable zero modes from bound states}\label{SecBoundStatesAndFactorizableStates}
In this section, we delve into the structure of the zero energy subspace. In Sec.~\ref{SecDefinitions}, we give generic definitions of two special subclasses of ZMs: bound states and factorizable ZMs. Bound states are entanglement-carrying many-body eigenstates that remain localized on a limited section of the lattice, while factorizable ZMs can be built as a collection of decoupled bound states. Both types of ZMs were first reported in the $U(1)$ $r=2$ East model~\cite{brighi2023}. Here, we provide the theoretical framework to understand these states and discuss a larger class of models that exhibit them. As an example, we analyze two classes of models that exhibit these types of states, the family of $U(1)$ East models, in Sec.~\ref{SecEast}, and the family of $U(1)$ East-West models, in Sec.~\ref{SecEastWest}.

\subsection{Definitions and generic construction}\label{SecDefinitions}
In order to define the notion of a bound state, we draw inspiration from compact localized states in the context of single particle physics, and generalize the concept to the many-body scenario (Sec.~\ref{SecBoundstates}). In the interacting case, the construction of these states becomes highly nontrivial, yet we show that these states can arise within degenerate subspaces and provide a sufficient condition for their existence. Then, we discuss how bound states can be used to construct factorizable eigenstates in systems with larger system sizes (Sec.~\ref{SecFactorizable}). Finally, we discuss the relationship between bound states and Hilbert space fragmentation (Sec.~\ref{SecFragmentation}).

\subsubsection{Bound states}\label{SecBoundstates}
Compact localized states (CLSs) have been extensively studied in the context of flat band systems~\cite{Flach2014a,Dias2015,Nandy2015,Pal2018,Ramachandran2017,MoralesInostroza2016,Mielke1991,Morfonios2021,Maimati2017,Maimati2021a,Maimati2021b,Maimati2021c,Xu2015,Sathe2021,Sathe2023,Rontgen2018}. These single particle eigenstates typically have a nonzero amplitude on a few nearby sites and a strictly vanishing amplitude on the rest of the system due to geometric frustration~\cite{Flach2014a,Maimati2017}. Then, the periodic structure of the lattice can create many copies of such states and, thus, a single particle flat band ~\cite{Rhim2019}. Regardless of the number of sites that a CLS occupies, its distinguishing feature is the following: it remains an eigenstate even when any number of additional sites are added to the lattice. Note that this is not the case for arbitrary eigenstates, which are not robust to the addition of lattice sites. This localization phenomenon arises due to one of two mechanisms that can be understood if one depicts the lattice sites and the particle hopping between them on a graph. Given an eigenstate with amplitude zero in some vertices (sites), adding a new vertex with an edge (hopping) to a vertex with amplitude zero will preserve the state as an eigenstate of the enlarged graph. If the new vertices are instead adjacent to vertices with a nonzero amplitude, the state can only remain an eigenstate through destructive interference at the new vertices.

Let us extend the notion of CLS to the many-body case. Now, the relevant graph is not the real space lattice, but the graph associated with the matrix of the many-body Hamiltonian, where each vertex represents a product state. Adding one site to the lattice will add many new vertices and edges to the many-body graph. First, let us propose a definition: 

\vspace{2mm}\noindent\textbf{Left and right bound states:} \textit{Let $|\psi^\ell \rangle$ be an eigenstate of a 1D many-body Hamiltonian on a lattice with $\ell$ sites that fulfills $\hat{H}^\ell |\psi^\ell \rangle=E|\psi^\ell \rangle$. Given the linear decomposition of $|\psi^\ell \rangle$ in the product state basis, $|\psi^\ell \rangle=\sum_{i}\alpha_i|s^\ell \rangle_i$, the state $|\psi^\ell \rangle$ is bound if and only if $\hat{H}^{\ell'}|\tilde{\psi}^{\ell'}\rangle=E|\tilde{\psi}^{\ell'}\rangle$ with $\ell'=\ell+q$ and $q\in\mathbb{N}$ where $|\tilde{\psi}^{\ell'}\rangle$ is
\begin{itemize}
    \item[-]  $|\tilde{\psi}^{\ell'}\rangle=\sum_i\alpha_i|k_{\rm L}\rangle^{\otimes q}\otimes|s^\ell \rangle_i$ (\textit{left bound}) or
    \item[-]  $|\tilde{\psi}^{\ell'}\rangle=\sum_i\alpha_i|s^\ell \rangle_i\otimes|k_{\rm R}\rangle^{\otimes q}$ (\textit{right bound})
\end{itemize}
and $|k_{\rm L(R)}\rangle$ are product states (padding) with support over $\ell_{\rm L}(\ell_{\rm R})$ sites.}\vspace{2mm}

The definition of a bound state directly follows from the above property: 

\vspace{2mm}\noindent\textbf{Bound state:} An eigenstate $|\psi^\ell \rangle$ of a 1D many-body Hamiltonian is a bound state if and only if it is simultaneously left and right bound. \vspace{2mm}

The simplest example of a bound state can naturally arise in the $U(1)$ East and East-West models when the padding states are both a single empty site, $|k_{\rm L}\rangle=|k_{\rm R}\rangle=|\circ\rangle$. Then, the bound state $|\psi^\ell \rangle$ lives within a given sector of the Hilbert space with a fixed particle number (see Secs. \ref{SecEast}, \ref{SecEastWest}). These kinds of bound states have a clear physical interpretation, they are collections of particles that remain bound in one section of the lattice. Thus, these bound states provide a direct analogy to the concept of CLSs in single particle systems. However, the rich structure of the many-body Hilbert space allows us to extend the notion of bound state to cases where the padding states are any product state configuration. Thus, more generally, bound states are eigenstates that remain stable upon expansion of the lattice by additional sites. Collective bound states are different from eigenstates in many-body-localized systems \cite{Chen2020}, whose localization stems from disorder and presents an exponential tail in Fock space instead of being compactly localized. Also, in contrast with other previously proposed notions of localization in the many-body space~\cite{danieli2020,danieli2021,danieli2022,vakulchyk2021,santos2019,santos2020,nicolau2023f} and nonergodic states~\cite{kuno2020,kuno2021,hart2020}, collective bound states do not rely on the presence of flat bands in the single-particle spectrum. 

The operation of adding a site to the lattice has a highly non-trivial effect in the many-body Hilbert space: take, for example, a spin-$1/2$ chain, adding one site doubles the dimension of the Hilbert space. As a consequence, typical eigenstates are not generally bound. We present a sufficient condition for the existence of a bound state by treating the Hamiltonian as an adjacency matrix with an associated graph. This condition imposes a special structure on the wave function, namely, that it is localized on a part of the graph that is not affected by enlarging the system size. 

\vspace{2mm}\noindent{\bf Sufficient conditions for a bound state:} Let $G^\ell $ be a graph with a set of vertices $\{|s^\ell \rangle_i\}$ associated to the matrix of $\hat{H}^\ell $ in the product state basis. Similarly, let $G^{\ell'}$ be the graph with vertices $\{|s^{\ell'}\rangle_i\}$ that corresponds to $\hat{H}^{\ell'}$, where $\ell'=\ell+n \ell_{\rm L}+m \ell_{\rm R}$ for $n,m\in\mathbb{N}$. Consider the following three conditions:
\begin{enumerate}[(i)]
    \item \textit{Recursivity:} There exists an induced subgraph\footnote{An induced subgraph of a graph $G^\ell $ is formed from a subset of vertices of $G^\ell $ and all edges in $G^\ell $ that start and end within that subset.} of $G^\ell $ with vertices $\{|s^\ell \rangle_{i,S}\}$ that is an induced subgraph of $G^{\ell'}$ with vertices $\{|k_{\rm L}\rangle^{\otimes n}\otimes|s^\ell \rangle_{i,S}\otimes|k_{\rm R}\rangle^{\otimes m}\}$. We denote this induced subgraph of $G^{\ell'}$ as $S^{\ell'}$, while $\bar{S}^{\ell'}$ denotes the induced subgraph which contains all vertices of $G^{\ell'}$ not included in $S^{\ell'}$.
    \item \textit{Sparse connectivity:} There is a subset of vertices $V_1$ of $S^{\ell'}$ that are not adjacent to any vertex in $\bar{S}^{\ell'}$. We denote the set of vertices that are adjacent to the vertices in $\bar{S}^{\ell'}$ as $V_2$.
    \item \textit{Compactness:} There exists an eigenstate $\hat{H}^\ell |\psi^\ell \rangle=E|\psi^\ell \rangle$, where $|\psi^\ell \rangle=\sum_{i\in V_1}\alpha_i|s^\ell \rangle_i$, i.e., the vertices in $S^{\ell'}$ adjacent to vertices in $\bar{S}^{\ell'}$ have amplitude zero.
\end{enumerate}
If these three conditions are met for any $n,m\in\mathbb{N}$, then the state $|\psi^\ell \rangle$ satisfies the definition of a bound state given above. However, for some models, if conditions (i-iii) are fulfilled for $n=m=1$, then they are fulfilled for any $n,m>1$, as we will discuss in Sec.~\ref{SecEast}. Intuitively, these conditions guarantee that the state $|\psi^\ell \rangle$ occupies the vertices of the graph that are not adjacent to the new vertices in the enlarged system. Thus, this state constitutes a bound state, in analogy with a single-particle CLS, without relying on destructive interference. Note that this is a sufficient but not necessary condition for a bound state to exist. In principle, one could find a bound state that has a nonzero amplitude on the vertices $V_2$ such that destructive interference makes the eigenstate robust to the increase in the lattice size. However, one expects these cases to be very rare in generic many-body quantum systems, as the addition of sites to the physical lattice results in an exponential increase in the order of the graph describing the Hilbert space and Hamiltonian. 

Let us discuss how conditions (i-iii) can be fulfilled in a physical system. Condition (i), recursivity, is guaranteed by the locality of the Hamiltonian: the increase in the size of the system might add or remove some vertices of $G^\ell $, but these alterations are limited by the range of the Hamiltonian terms. Thus, for a large enough system size, one can always find an induced subgraph $S^{\ell'}$. Condition (ii), sparse connectivity, can either be guaranteed by $U(1)$ symmetry, not all vertices in $S^{\ell'}$ can be coupled to a vertex in $\bar{S}^{\ell'}$ due to particle conservation, or by the presence of kinetic constraints, which remove some edges in the graph. Finally, condition (iii) is the most nontrivial to be fulfilled, as generally we expect that the wave functions of eigenstates have nonzero amplitudes on nearly all computational basis states (for generic interacting models). Below, we show that for the considered models, condition (iii) can be fulfilled \emph{within} the degenerate subspace of ZMs induced by chiral symmetry.  

Given a set of $C$ degenerate eigenstates $\{|\psi_i\rangle\}$, any superposition $|\tilde{\psi}_i\rangle=\sum_{i=1}^C\alpha_i|\psi_i\rangle$ is also an eigenstate of the system. Therefore, one can search for a rotation of the degenerate subspace that yields eigenstates that satisfy the compactness condition (iii), and thus have amplitude zero on the set of vertices $V_2$. In a particular model, bound states can be identified by constructing an operator $\hat{O}$ that discriminates between the sets of vertices $V_2$ and $V_1$. The operator $\hat{O}$ must be positive semidefinite and give $\hat{O}|\psi\rangle=0$ for those eigenstates that only have support on the vertices in $V_1$, $|\psi\rangle=\sum_{i\in V_1}\alpha_i|s_i\rangle$, and give $\langle\psi|\hat{O}|\psi\rangle>0$ otherwise. Numerically, one can find such states by resolving the operator $\hat{O}$ in the degenerate subspace, by defining a matrix $\Theta({\hat{O}})$ with matrix elements
\begin{equation}\label{EqKernelOperator}
    \Theta_{ij}({\hat{O}})=\left\langle\psi_i\right|\hat{O}\left|\psi_j\right\rangle.
\end{equation}
Then, the kernel of the matrix $\Theta({\hat{O}})$ will contain the amplitudes $\alpha_i$ to construct the bound states from the set of states $\{|\psi_i\rangle\}$. In chiral systems, the degenerate subspace appears at zero energy, such that if condition (iii) is fulfilled, it leads to a bound state that is a ZM. 

\subsubsection{Factorizable eigenstates}\label{SecFactorizable}

Bound states are the building blocks of the second subclass of ZMs that we will discuss, factorizable ZMs, which are built from ZMs of smaller system sizes. Let us first consider a more general definition. 

\vspace{2mm}\textbf{Factorizable eigenstate}: \textit{An eigenstate $|\psi^{\ell}\rangle$ that can be decomposed as the tensor product of $m$ eigenstates of smaller system sizes $\ell_i$ (building states) separated by strings of product states $|K^d\rangle$ with support over $d_i$ sites (decoupling states),
\begin{equation}\label{EqFactorizable}
|\psi^{\ell}\rangle=|\psi^{\ell_1}\rangle\otimes|K^{d_1}\rangle\otimes|\psi^{\ell_2}\rangle\otimes\cdots\otimes|\psi^{\ell_m}\rangle,
\end{equation}
where the states used as building blocks fulfill $\hat{H}^{\ell_i}|\psi^{\ell_i}\rangle=E^{\ell_i}|\psi^{\ell_i}\rangle$ and $m\in\mathbb{N}$ with $m\geq2$.}\vspace{2mm}

\noindent As the energy of $|\psi^{\ell}\rangle$ is given by $E^\ell =\sum_iE^{\ell_i}$, if all the building states in Eq.~(\ref{EqFactorizable}) are ZMs, $E^{\ell_i}=0$, then, the resulting state is also a ZM, $E^\ell=0$. We call these states \textit{factorizable ZMs}. 

Note that an arbitrary construction of the form \eqref{EqFactorizable} does not necessarily yield an eigenstate. A possible way to build such a state is to fulfill the following three conditions:
(I) The building states $|\psi^{\ell_i}\rangle$ with $i=2,3,...,m-1$ are bound states, while $|\psi^{\ell_1}\rangle$ can be just right bound and $|\psi^{\ell_m}\rangle$ can be just left bound. 
(II) The decoupling product states $|K^d\rangle$ must be constructed from the states $|k_{\rm R}\rangle$ and $|k_{\rm L}\rangle$, which leave the building states invariant, and thus make them bound states.
(III) The decoupling product states $|K^d\rangle$ must be dynamically inert and have a support $d$ large enough to ensure that the building states are not directly coupled through interactions.

For example, assuming the padding states are the same for $|\psi^{\ell_1}\rangle$ and $|\psi^{\ell_2}\rangle$, one can use $|K^d\rangle=|k_{\rm R}\rangle\otimes|k_{\rm L}\rangle$ as a decoupling configuration. The simplest case arises in the $U(1)$ East and East-West models, where $|k_{\rm L}\rangle=|k_{\rm R}\rangle=|\circ\rangle$, such that $|K^d\rangle$ simplifies to a string of empty sites $|K^d\rangle=|\circ\rangle^{\otimes d}$. If conditions (I-III) above are fulfilled, then $|\psi^\ell \rangle$ is a factorizable eigenstate. These states will generally present a characteristic feature in their bipartite Von Neumann entanglement entropy, $\mathcal{S}=-\operatorname{tr}(\rho_A \ln \rho_A)$, where $\rho_A$ is the reduced density matrix of subsystem $A$. If one represents $\mathcal{S}$ as a function of the last site of subsystem $A$, the entanglement presents a dip to zero entanglement for each string of empty sites in the factorizable eigenstate. Such a structure is unexpected for highly excited states, which are usually strongly entangled.\\

\subsubsection{Generalizations and relation to Hilbert space fragmentation}\label{SecFragmentation}

The notions of bound and factorizable eigenstates that we have proposed above can be easily extended to other systems. For example, we give a definition of bound states for 1D lattices, which sharply defines a left and a right boundary and the derived notions of left and right bound states. However, one could expand this definition to $d$-dimensional geometries, such that an eigenstate would be completely bound if it remains robust to the addition of product state configurations along its $(d-1)$-dimensional edge. Similar to the 1D case, one could reduce the problem to its many-body graph representation to find an analogous sufficient condition for the existence of bound states. Also, while the present definition of 1D factorizable eigenstates assumes open boundary conditions, one can generalize it to periodic boundary conditions by simply adding an additional decoupling state between the first and last building states. One might also consider a generalized notion of bound state that is robust to the addition of any state, including entangled states. In this case, the localization mechanism of the eigenstate in the many-body graph would only be apparent in a suitable entangled basis.

Finally, the presence of bound states in a many-body Hamiltonian can also be related to Hilbert space fragmentation. Take, for example, a Hamiltonian $\hat{H}^\ell$ on $\ell$ sites with an associated graph $G^\ell$ such that all its eigenstates are bound states. As this set of bound states forms a complete basis, all vertices in $G^\ell$ must have weight on at least one of these states. However, the compactness condition (iii) requires bound states to only occupy the vertices of the graph that do not acquire any new edges in the enlarged graph $G^{\ell'}$ (see Sec.~\ref{SecBoundstates}). Thus, for compactness (iii) to be fulfilled for all eigenstates, no vertices in $G^\ell$ can acquire new edges in $G^{\ell'}$. That is, according to the sufficient condition presented above, $G^\ell$ will in general be a connected component of $G^{\ell'}$ such that $\hat{H}^{\ell'}$ takes a block diagonal form in the product state basis. If the number of sectors grows exponentially with system size (e.g., as in the $U(1)$ East model), there is classical Hilbert space fragmentation. This fragmentation is classical because this set of bound states admits a product state basis, as they comprise all eigenstates of $\hat{H}^\ell$. Also, these states are trivially bound because the connected graph component that hosts them does not change when adding more sites. The presence of nontrivial bound states that do not admit a product state basis can also lead to quantum Hilbert space fragmentation, which is only readily apparent in an entangled basis. We discuss an example of quantum fragmentation in the $U(1)$ East model in Sec.~\ref{SecEast}.

\begin{figure*}[t]
    \includegraphics[width=2\columnwidth]{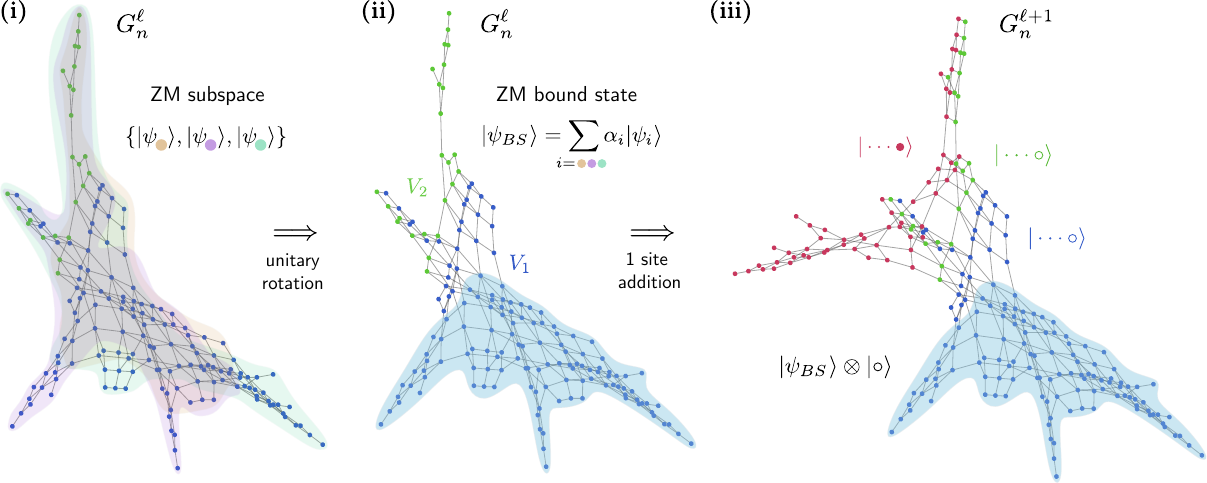}
    \caption{Illustration of the procedure to construct a bound state within the ZM subspace of the $U(1)$ East model restricted to the largest classical fragmentation sector. (i) Consider a set of ZM $\{|\psi\rangle\}$ on a graph $G_n^\ell $ of a lattice with $\ell$ sites and $n$ particles. (ii) Rotate the subspace to construct a ZM localized on the set of vertices $V_1$ (blue vertices), $|\psi_{BS}\rangle$. (iii) Add one site to the right boundary of the lattice, which yields the graph $G_n^{\ell+1}$. Note that $G_n^{\ell+1}$ contains $G_n^\ell $ as an induced subgraph and also an extra set of vertices of the form $|\cdots\bullet\rangle$ (red vertices). As the red vertices are only connected to the vertices in the set $V_2$, the state $|\psi_{BS}\rangle\otimes|\circ\rangle$ remains an eigenstate of the model, and thus, $|\psi_{BS}\rangle$ is a right bound eigenstate. For the $U(1)$ East model, adding one site to the left also yields the same graph, $G_n^{\ell+1}=G_n^{\ell+2}$, as no particle can reach that site, and thus, $|\psi_{BS}\rangle$ is not only right bound but is a full bound state.}
    \label{FigSubspaceRotation}
\end{figure*}

\subsection{East Model}\label{SecEast}
In this section, we analyze the $U(1)$ East model as an example of a local and chiral Hamiltonian that exhibits a large degenerate subspace. We discuss how this model fulfills the existence conditions (i-iii) for bound states (see Sec.~\ref{SecBoundstates}) and explain how to find them numerically within the ZM subspace. Then, we show how one can construct factorizable eigenstates and discuss how these affect the whole energy spectrum. Finally, we discuss how both classical and quantum fragmentation in this model can be understood in terms of the boundedness of its eigenstates.

\subsubsection{Bound state construction}
The kinetic constraint in the $U(1)$ East model induces a pattern of classical fragmentation where the largest sector can be generated from a domain wall state of the form $|\bullet\rangle^{\otimes N}\otimes|\circ\rangle^{\otimes(L-N)}$. In this sector, the system has a natural particle filling determined by $L=(r+1)N-r$, which gives the size of the most diluted state $(|\bullet\rangle\otimes|\circ\rangle^{\otimes r})^{\otimes(N-1)}|\bullet\rangle$. For example, for range $r=2$ and $N=3$, the domain wall state $|{\bullet\bullet\bullet\circ\circ\circ\circ}\rangle$ can expand to the most diluted state $|\bullet\circ\circ\bullet\circ\circ\bullet{\rangle}$. Any additional sites do not change the dynamics of the system, as the particles cannot reach them, while fewer sites would make the boundaries affect the dynamics. All other sectors present frozen regions composed of empty sites, such that their dynamics can be understood from those of smaller system sizes. Thus, in what follows, we focus on the largest sector. 

For $U(1)$ conserving systems, when the Hilbert space is enlarged by adding sites to the lattice, bound states can arise within a fixed sector of the $U(1)$ charge. For the $U(1)$ East model, these states emerge for a given particle number, which makes them robust to the addition of empty sites. Here, we analyze the largest components of the graphs $G_n^\ell$ and $G_n^{\ell+1}$, where $n$ is the fixed particle number and $\ell$ is the number of sites, and fix both padding states to $|k_{\rm L(R)}\rangle=|\circ\rangle$. Figure~\ref{FigSubspaceRotation} illustrates the conditions and procedure for the construction of ZM bound states discussed in the previous section. Figs.~\ref{FigSubspaceRotation}(i,ii) show an example of the largest connected component of the graph $G_n^\ell$, while Fig.~\ref{FigSubspaceRotation}(iii) shows the graph corresponding to the lattice with one additional site and the same particle number, $G_n^{\ell+1}$. The sets of vertices $V_1$ and $V_2$ that determine the sparsity condition (ii) are indicated in Fig.~\ref{FigSubspaceRotation}(ii). 

In this case, the recursivity condition (i) is fulfilled because the induced subgraph of $G_n^{\ell'}$ contains all vertices in $G_n^\ell$, i.e., it is the whole graph. This occurs because adding empty sites does not constrain the dynamics or enhance the mobility of particles in $G_n^\ell$. This is illustrated in Fig.~\ref{FigSubspaceRotation}(iii), where the new vertices (in red) are those that have the last site full, while the ones with the last site empty (in blue and green) form the induced subgraph given by $G_n^{\ell}$ in Fig.~\ref{FigSubspaceRotation}(i,ii). The sparse connectivity condition (ii) is guaranteed by particle conservation ($U(1)$ symmetry). The range of the kinetic constraint $r$ determines a small set of vertices $V_2$ (green) where a particle can jump to the last site, making them adjacent to the new vertices (red), while the vertices in $V_1$ (blue) are not adjacent [see Fig.~\ref{FigSubspaceRotation}(ii-iii)]. A possible operator that distinguishes between the two sets of vertices is $\hat{O}=\hat{n}_L\sum_{j=1}^{r}\hat{n}_{L-j}$, which can then be used to search for ZMs that fulfill the compactness condition (iii), i.e., ZMs that have amplitude zero in the vertices in $V_2$. These states are constructed from the kernel of the matrix $\Theta({\hat{O}})$ defined in Eq.~\eqref{EqKernelOperator}. For example, for $r=2$, the operator can take the form $\hat{O}=\hat{n}_{L-1}\hat{n}_L+\hat{n}_{L-2}\hat{n}_L$. 

In summary, the procedure to generate a bound state is the following. Consider a ZM subspace with three hypothetical states that live in $G_n^\ell$, whose area of support is depicted qualitatively in Fig.~\ref{FigSubspaceRotation}(i). We perform a unitary rotation using the operator $\hat{O}$ that yields a ZM, $|\psi_{BS}\rangle$, that has support only on the set of vertices $V_1$ [Fig.~\ref{FigSubspaceRotation}(ii)]. When adding an additional site to the lattice, $|\psi_{BS}\rangle\otimes|\circ\rangle$ remains an eigenstate, as it has no support on $V_2$ (green vertices), and is thus unaffected by the new vertices (red) [Fig.~\ref{FigSubspaceRotation}(iii)]. As this model features nearest neighbor hopping, when one adds a second site to the right boundary, only some red vertices will be adjacent to the new vertices in $G_n^{\ell+2}$, such that $|\psi_{BS}\rangle\otimes|\circ\circ\rangle$ will also remain an eigenstate. One can repeat this process iteratively, which shows that the state $|\psi_{BS}\rangle$ is right bound, as it is robust to the addition of an arbitrary number of empty sites to its right. Additionally, adding empty sites to the left boundary leaves the graph invariant, as the leftmost particle is always frozen, which also makes $|\psi_{BS}\rangle$ left bound, and thus, a bound state. 

\subsubsection{Factorizable zero modes}

In order to build factorizable eigenstates, we also need a suitable decoupling state, which in this case is composed of a string of empty sites. The minimum number of empty sites needed to prevent coupling between adjacent bound states is $r+1$, where $r$ is the range of the kinetic constraint. For example, for $r=2$ and a state of the form $|\psi_{n_1}^{\ell_1}\rangle\otimes|{\circ\circ\circ}\rangle\otimes|\psi_{n_2}^{\ell_2}\rangle$, the leftmost particle in the bound state $|\psi_{n_2}^{\ell_2}\rangle$ does not interact with the rightmost particle in the bound state $|\psi_{n_1}^{\ell_1}\rangle$. In order to find these states numerically, one can use the operator $\hat{\mathcal{W}}_j=\sum_{t=0}^{r}\hat{n}_{j+t}$ to search for inert configurations at different sites $j$ of the lattice. In analogy with bound states, the kernel of matrix $\Theta(\hat{\mathcal{W}}_j)$, defined in Eq.~\eqref{EqKernelOperator}, will give the factorizable eigenstates, while the total number will be the number of linearly independent solutions obtained through this method. 

\begin{figure}[t]
    \includegraphics[width=\columnwidth]{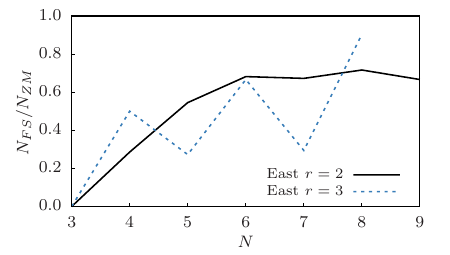}
    \caption{Ratio of factorizable ZMs to the total number of ZMs as a function of the number of particles $N$ for the largest sector of the $U(1)$ East model and $r=2,3$. The proportion of factorizable ZMs increases with $N$, reaching a high saturation value for $r=2$ and exhibiting a modulated behavior for $r=3$. As the ZM subspace grows exponentially with $N$, any non-decreasing ratio $N_{FS}/N_{ZM}$ suggests that the number of factorizable ZMs also increases exponentially with $N$.}
    \label{FigRatio}
\end{figure}

Figure \ref{FigRatio} shows the fraction of factorizable ZMs to the total number of ZMs as a function of the number of particles, $N$, for the largest sector and $r=2,3$. For both ranges, the fraction rapidly increases with $N$; $r=2$ seems to reach a saturation value $\sim 0.7$ while $r=3$ exhibits an even-odd modulated behavior and reaches $\sim 0.9$ for $N=8$. There is always a remaining fraction of ZMs that cannot be decomposed into smaller ZM bound states. Appendix~\ref{SecAppendix} contains supplementary data on the number of ZMs and ZM factorizable eigenstates for the family of $U(1)$ East models. These results reveal the high degree of structure of the ZM subspace: at each system size, a large proportion of the ZMs can be constructed from ZM bound states from smaller system sizes. Additionally, if these factorizable states are also bound, they can in turn become the building blocks for more complex factorizable states when increasing the system size further. 

\begin{figure}[t]
    \includegraphics[width=\columnwidth]{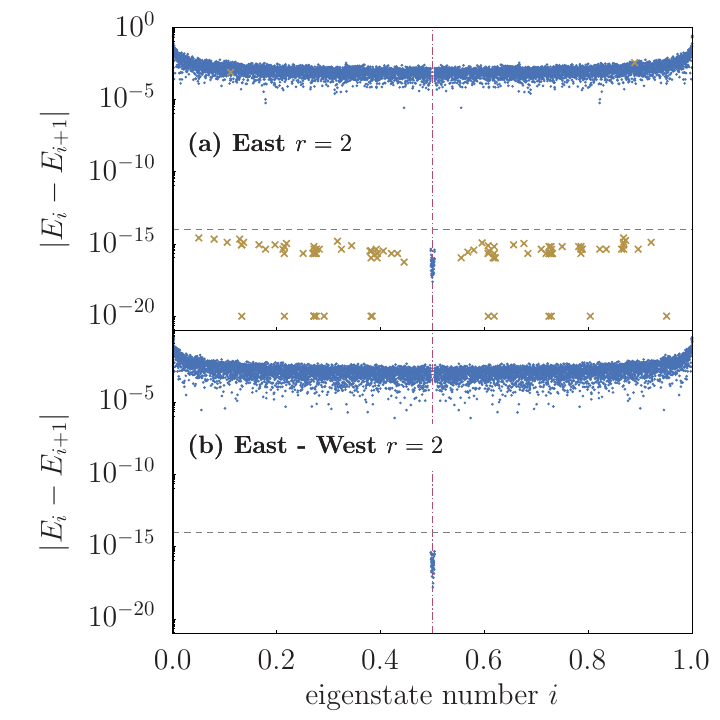}
    \caption{Comparison of the energy differences $|E_i-E_{i+1}|$ between the (a) $U(1)$ East and the (b) $U(1)$ East-West models. The factorizable eigenstates with nonzero energy are highlighted by yellow crosses, and the exact degeneracies are fixed at $10^{-20}$ for visibility. The stronger kinetic constraints in the East model allow ZM bound states to affect the whole energy spectrum, giving rise to degenerate eigenenergies away from $E=0$ due to factorizable eigenstates. (a) Largest sector for $r=2$, $N=7$ and $L=19$ and (b) full Hilbert space for $r=2$, $N=6$ and $L=16$.
    }\label{FigDegeneracy}
\end{figure}

The $U(1)$ East model presents a special feature: all eigenstates are left bound states due to the kinetic constraint, as any leftmost particle with only empty sites on its left is always frozen. As a result, the rightmost building state in Eq.~\eqref{EqFactorizable} can be an arbitrary eigenstate, as long as it has the appropriate number of sites and particles. The resulting state is a factorizable eigenstate that can have nonzero energy, which leads to the appearance of additional degeneracies besides the ZM subspace. Figure \ref{FigDegeneracy}(a) shows the energy differences $E_{i+1}-E_i$ for $r=2$ and $N=7$ with the factorizable eigenstates with nonzero energy indicated by yellow crosses and the exact degeneracies fixed at $10^{-20}$ for visibility. For $N=[3-7]$, all extra degeneracies are due to the presence of factorizable eigenstates, behavior that we expect to hold for arbitrary~$N$. At $N=7$, only a pair of factorizable eigenstates do not come from a degenerate subspace. 

\subsubsection{Classical and quantum fragmentation}
The family of $U(1)$ East models presents classical and quantum Hilbert space fragmentation due to the presence of bound states and factorizable eigenstates. The first mechanism of classical fragmentation, the localization of the leftmost particle, appears due to the fact that all eigenstates for a given system size are left bound. The second mechanism arises from the largest sector of a sufficiently dilute Hamiltonian $\hat{H}_n^\ell$ such that $\ell\geq(r+1)n-r$. As the particles cannot spread further to the right, all eigenstates of $\hat{H}_n^\ell$ are bound states, and one can form dynamically disconnected sectors by padding a bound state from a dilute region with $r+1$ empty sites such that the next particle on the right becomes frozen.  

As discussed above, the $U(1)$ East models present additional bound states in the ZM subspace. Those do not admit a product state basis, and thus the resulting fragmentation is quantum. Consider for example, the following set of states for $r=2$~\cite{brighi2023}, $|\psi_{BS}^{\ell_1}\rangle\otimes|{\circ \circ\circ}\rangle \otimes|\psi^{\ell_2}\rangle,$ which is composed of a ZM bound state (e.g., $|{\bullet \bullet \circ \circ \bullet }\rangle-|{\bullet \circ \bullet \bullet \circ }\rangle/\sqrt{2}$) separated by three empty sites from an arbitrary eigenstate on $\ell_2$ sites on the right. Each bound state generates a dynamically disconnected sector with nontrivial dynamics due to the arbitrary eigenstates on the right. As the set of bound states on $\ell_1$ sites, $\{|\psi_{BS}^{\ell_1}\rangle\}$, does not support a product state basis, the resulting fragmentation is quantum, as it is only apparent in a suitable entangled basis.

\subsubsection{Dynamical signatures}
\begin{figure}[t]
    \includegraphics[width=\columnwidth]{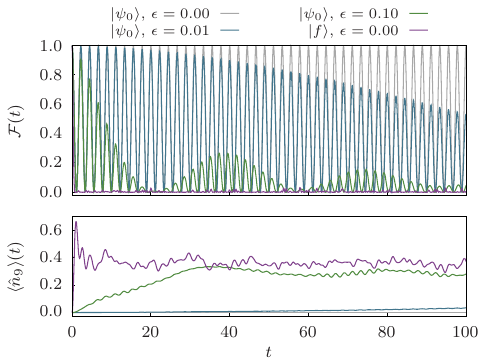}
    \caption{Dynamical signature of bound states and quantum Hilbert space fragmentation in the $U(1)$ East model for $N=6$ particles. Top panel: time evolution of the fidelity of an initial state with support on a bound state, $|\psi_0\rangle$ [Eq.~(\ref{EqInitialState})], and a random product state $|f\rangle$. The fidelity revivals are quite robust to the introduction of an uncorrelated hopping term with perturbation strength $\epsilon$. The bottom panel shows that the damping of fidelity revivals correlates with the particle leakage through the padding sites, as represented by the expectation value $\langle \hat{n}_9\rangle$.}
    \label{FigUncorrelated}
\end{figure}
In this section, we briefly examine the consequences of the presence of bound states on the dynamics of the $U(1)$ East model. We consider the time evolution of the fidelity $\mathcal{F}(t)=|\langle \psi_0|\psi(t)\rangle|^2$ of a certain initial state $|\psi_0\rangle$ with weight on a bound state and partial overlap with a factorizable state. The initial state is given by the tensor product
\begin{equation}\label{EqInitialState}
    |\psi_0\rangle=|\psi_{BS}\rangle\otimes|{\circ\circ\circ}\rangle\otimes|{\bullet\circ\bullet\circ\circ}\rangle,
\end{equation}
where $|\psi_{BS}\rangle$ is a bound state with four particles,
\begin{equation}
    \begin{aligned} |\psi_{BS}\rangle=& \frac{1}{2}(|{\bullet\bullet\circ\circ\bullet\circ\circ \bullet } \rangle-|{\bullet \bullet \bullet \circ \circ \circ \circ \bullet}\rangle)\\
    &+\frac{1}{4}(|{\bullet \circ \circ \bullet \bullet \bullet \circ \circ }\rangle+|{\bullet \circ \bullet \bullet \circ \bullet \circ \circ }\rangle \\ & +|{\bullet \circ \circ \bullet \circ \bullet \bullet \circ }\rangle+|{\bullet \bullet \bullet \circ \circ \bullet \circ \circ }\rangle\\&-|{\bullet \bullet \circ \bullet \bullet \circ \circ \circ }\rangle-|{\bullet \bullet \circ \circ \bullet \bullet \circ \circ }\rangle \\ & -|{\bullet \circ \circ \bullet \bullet \circ \circ \bullet }\rangle-|{\bullet \circ \bullet \circ \bullet \circ \bullet \circ }\rangle),\end{aligned}
\end{equation}
and the right part is a product state that includes three empty sites that implement the padding that decouples the bound state from the particles on the right. Figure \ref{FigUncorrelated} depicts the evolution of the fidelity of this initial state compared to that of a random product state $|f\rangle$. The fidelity of $|\psi_0\rangle$ exhibits perfect revivals (solid gray line), while the fidelity of the random product state decays rapidly to zero and fluctuates near that value (solid purple line). These revivals are a direct consequence of the presence of bound states and the quantum Hilbert space fragmentation they generate: the left part of the state is completely frozen such that the dynamics are completely determined by the two particles on the right. As the dynamics only depend on the right part of the state, analogous revivals can be obtained by introducing other bound states on the left, regardless of the number of physical sites or Fock states they occupy (a similar case is analyzed in detail in \cite{brighi2023}).  

In order to check the robustness of the bound states, we introduce a perturbation to the model, namely, we add a term to the Hamiltonian that implements uncorrelated hopping with strength $\epsilon$, $\delta\hat{H}=\epsilon\sum_{i=1}^{L-1}(\hat{c}_{i+1}^{\dagger} \hat{c}_i+\text {h.c.})$. This term destroys both classical and quantum Hilbert space fragmentation by adding hopping terms to the Hamiltonian which break the kinetic constraints of the $U(1)$ East model. As a result, the degeneracy of the ZM subspace is partially lifted as the chiral mismatch lower bound [Eq.~(\ref{EqMismatch})] becomes saturated (see discussion in Appendix \ref{SecAppendix2}). However, the effects of the fragmentation structure in the unperturbed case persist up to relatively strong perturbations. For the initial state $|\psi_0\rangle$ and different perturbation strengths $\epsilon=\{0.1,0.01\}$, Fig.~\ref{FigUncorrelated} shows the evolution of the fidelity in the perturbed model (top panel). Even for relatively strong perturbations, $\epsilon\sim0.1$, that are much larger than level spacing, the state exhibits strong revivals which indicate a long term memory of the initial state that is not present for the random state in the absence of perturbations ($\epsilon=0$). This indicates that the bound states are quite robust to the addition of uncorrelated hopping terms. 

The damping of the oscillations for $\epsilon>0$ can be directly related to the leakage through the padding sites, which can be observed in the expectation value $\langle \hat{n}_{9}\rangle$, the first empty site of the padding (bottom panel of Fig.~\ref{FigUncorrelated}). For $\epsilon=0.01$, the oscillations are barely damped, as $\langle \hat{n}_{9}\rangle$ barely grows and stays below 0.04, while for $\epsilon=0.1$, the revivals decay more rapidly, as $\langle \hat{n}_{9}\rangle$ grows much quickly, approaching the random state value. In contrast, the leakage is exactly zero at all times for the state $|\psi_0\rangle$ and $\epsilon=0$, denoting quantum fragmentation. In Appendix \ref{SecAppendix2}, we examine the consequences of another type of disorder, hopping disorder, which preserves a high lower bound on the number of bound states. In Sec.~\ref{SecGeneralization}, we discuss further dynamical signatures of bound states in another model, the $U(1)$ North-East model.

\subsection{East-West Model}\label{SecEastWest}
In this section, we analyze the family of $U(1)$ East-West models, which present an additional kinetic term in the Hamiltonian that restores inversion symmetry. While these models also present bound states and factorizable ZMs, inversion symmetry makes them different than those found in the $U(1)$ East models. First, we find an analytical description of bound states and describe the generation of factorizable ZMs. Then, we discuss the effect of these states on the energy spectrum and the structure of the Hilbert space.

\subsubsection{Bound states}
\begin{figure}[t]
    \includegraphics[width=\columnwidth]{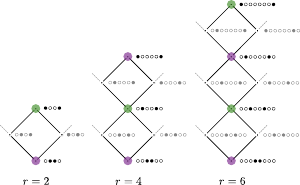}
    \caption{Representation of the two-particle ZM bound states for the $U(1)$ East-West model for an arbitrary even range $r=2,4,6,...$, where the circle radii indicate the amplitude and the color (green or purple) indicates the phase ($\pi$ or $0$, respectively). This construction arises from destructive interference induced by the inversion symmetry of the model, and it can be generalized to an arbitrary number of particles, forming the building blocks to construct arbitrarily large factorizable ZMs.}
    \label{FigBoundstate}
\end{figure}

The inversion symmetry of the $U(1)$ East-West models can be used to analytically construct a ZM bound state for arbitrary $N$ and $r$. Let us first consider the case of two particles separated by $r$ empty sites, for example, $|\bullet\circ\circ\bullet\rangle$ for $r=2$, and assume that additional empty sites surround this pair (not depicted here). Moving the particles closer together generates the product states $|\circ\bullet\circ\bullet\rangle$, $|\bullet\circ\bullet\circ\rangle$, and $|\circ\bullet\bullet\circ\rangle$, which form a symmetric diamond structure where only the inversion asymmetric states, $|\circ\bullet\circ\bullet\rangle$ and $|\bullet\circ\bullet\circ\rangle$, are adjacent to the other vertices of the graph. This structure supports a ZM of the form ($|\bullet\circ\circ\bullet\rangle-|\circ\bullet\bullet\circ\rangle)/\sqrt{2}$ that arises due to the destructive interference at the inversion asymmetric states, $|\circ\bullet\circ\bullet\rangle$ and $|\bullet\circ\bullet\circ\rangle$. One can generalize this ZM construction for an arbitrary even range $r$ by populating only the inversion symmetric states with the same amplitude and alternating phases, $\sum_{j=0}^{r/2}(-1)^j|\circ\rangle^{\otimes j}|\bullet\rangle|\circ\rangle^{\otimes (r-2j)}|\bullet\rangle|\circ\rangle^{\otimes j}$. For example, for $r=4$, the state is $(|\bullet\circ\circ\circ\circ\bullet\rangle-|\circ\bullet\circ\circ\bullet\circ\rangle+|\circ\circ\bullet\bullet\circ\circ\rangle)/\sqrt{3}$. We represent these ZMs for $r=2,4,6$ in Fig.~\ref{FigBoundstate}, which illustrates how this pattern of destructive interference is a direct result of the inversion symmetry of the model. Note that this construction is not possible for odd $r$, as it requires a symmetric compressed state of the form $|\circ\rangle^{\otimes r/2}|\bullet\rangle|\bullet\rangle|\circ\rangle^{\otimes r/2}$. 

This ZM construction can be further generalized to an arbitrary number of particles by using the most diluted configuration, $(|\bullet\rangle\otimes|\circ\rangle^{\otimes r})^{\otimes(N-1)}|\bullet\rangle$, as a seed state. For example, for $N=3$ and $r=2$, the seed state is $|{\bullet\circ\circ\bullet\circ\circ\bullet}\rangle$. The product states that are populated can be obtained by taking each local particle pair in the seed state, $\cdots|\bullet\rangle|\circ\rangle^{\otimes r}|\bullet\rangle\cdots$, and moving both particles closer together such that the state remains symmetric with respect to the center of the particle pair. These two-particle swaps eventually yield the most compressed state and can be performed for each particle pair independently, which generates many configurations. The amplitude will be constant for all these product states, while the phase will be given by the parity of the minimum number of two-particle swaps required to reach the diluted configuration. For example, for $r=2$ and three particles, the phases of the product states are the following:  $|\bullet\circ\circ\bullet\circ\circ\bullet\circ\circ\bullet\rangle$ (zero particle swaps, phase $0$);  $|\bullet\circ\circ\circ\bullet\bullet\circ\circ\circ\bullet\rangle$, $|\circ\bullet\bullet\circ\circ\circ\bullet\circ\circ\bullet\rangle$, and $|\bullet\circ\circ\bullet\circ\circ\circ\bullet\bullet\circ\rangle$ (1 particle swap, phase $\pi$); $|\circ\bullet\bullet\circ\circ\circ\circ\bullet\bullet\circ\rangle$ (2 particle swaps, phase $0$). The resulting state can be generated by a matrix product operator with bond dimension $\chi=3$, 
\begin{align}
    \hat{\mathcal{F}}=& \prod_{i=1}^{\frac{L-1}{r+1}}\left(1-\sum_{j=1}^{r/2}(-1)^j\hat{\mathcal{F}}_{(r+1)i-r}^j\right)\\
    \hat{\mathcal{F}}_i^j=&\hat{c}_i\hat{c}_{i+j}^\dagger \hat{c}_{i+r+1-j}^\dagger \hat{c}_{i+r+1}+\rm{h.c.}
\end{align}
acting on the most diluted state $(|\bullet\rangle\otimes|\circ\rangle^{\otimes r})^{\otimes(N-1)}|\bullet\rangle$, where $\hat{\mathcal{F}}_i^j$ acts on the sites $\{i,i+r+1\}$. The operator can be expressed in matrix product form as $\hat{\mathcal{F}}=\langle L|\prod_{i=0}^{(L-1)/(r+1)} M_{(r+1)i+1}|R\rangle$, where
\begin{equation}\label{EqMPOMatrix}
	M_i=\begin{pmatrix}
		1 & \hat{A}^+ & \hat{A}^- \\[1mm]
		\hat{\sigma}_{i}^+ & 0 & 0\\[1mm]
		\hat{\sigma}_{i}^- & 0 & 0\\
	\end{pmatrix},
\end{equation}
with
\begin{equation}
    \hat{A}^+= \hat{\sigma}_{i}^+\sum_{n=1}^{r/2}(-1)^n\hat{\sigma}_{i+n}^-\hat{\sigma}_{i+r-n+1}^-,
\end{equation}
and $\hat{A}^-=\hat{A}^{+\dagger}$. The left and right vectors are $|L\rangle=|R\rangle=(1;0;0)$. 

Here, we have discussed open boundary conditions. However, as particle pairs can travel indefinitely in the $U(1)$ East-West model, periodic boundary conditions are also a natural option to consider. In this case, one could also construct ZM bound states by introducing additional empty sites to the seed state to separate the first and last particles, $(|\bullet\rangle\otimes|\circ\rangle^{\otimes r})^{\otimes N}$.

\subsubsection{Factorizable zero modes}
The construction of factorizable eigenstates from these bound states is similar to the $U(1)$ East case: the decoupling states are strings of empty sites given by the range of the model, $|\circ\rangle^{\otimes (r+1)}$, and they can be found numerically as described in the previous section. However, there is an important difference in the East-West models: arbitrary eigenstates are not left bound, as both the right and the left neighbors of a given particle can enable hopping. Thus, all building states of a factorizable eigenstate must be bound states, with the exception of the leftmost and rightmost building states with open boundary conditions, which can be merely right and left bound states, respectively. As a consequence, there are no additional degeneracies besides the ZM subspace, where all the factorizable eigenstates reside (see Fig.~\ref{FigDegeneracy}(b) for $r=2$, $N=6$, and $L=16$). In Table~\ref{TabWest} of Appendix \ref{SecAppendix}, we summarize the number of ZM, ZM bound states, and factorizable ZMs for different numbers of particles and increasing system size. Both types of ZMs, bound states and factorizable eigenstates, exhibit an approximately exponential growth in their number with the size of the system. 

In the $U(1)$ East-West model, any state where all particles are separated by $r+1$ or more empty sites is a bound state and also a frozen state, as these are product states that are dynamically disconnected from all other states. However, as there is no extensive number of bound states, as is the case in the East model, there is no classical fragmentation. In addition to these frozen states, there are non-trivial ZM bound states generated by Eq.~\eqref{EqMPOMatrix} that live in the connected component of the many-body graph and can be used to generate factorizable ZMs. Consider for example the subspace of factorizable states $|\psi_{BS}^{\ell_1}\rangle\otimes|\circ\rangle^{\otimes q}\otimes|\psi^{\ell_2}\rangle$, where the left state is a particular bound state on $\ell_1$ sites and the right state $|\psi^{\ell_2}\rangle$ is an arbitrary bound state on $\ell_2$ sites. As the states on the right $|\psi^{\ell_2}\rangle$ can only be composed of ZM bound states, which are degenerate, the unitary rotation into this subspace does not lead to quantum fragmentation, with a sector with nontrivial dynamics, but instead yields a set of frozen states with energy zero. That is, the Hamiltonian matrix in this basis becomes partially diagonalized as it presents a diagonal block of ZMs.

\subsection{Generalization}\label{SecGeneralization}
In this section, we show that bound states are not a particular feature of the $U(1)$ East and East-West models by briefly describing two additional models with distinct properties. First, we discuss a two-dimensional (2D) generalization of the $U(1)$ East model, the $U(1)$ North-East model, which hosts 2D bound states. Secondly, we discuss the pair-flip model, an inversion symmetric model that lacks a  $U(1)$ symmetry related to particle-conservation, which hosts bound states both within the zero mode subspace and at nonzero energies. These examples illustrate the generality of our bound state construction discussed above.

\subsubsection{North-East model}
\begin{figure}[tb]
    \includegraphics[width=\columnwidth]{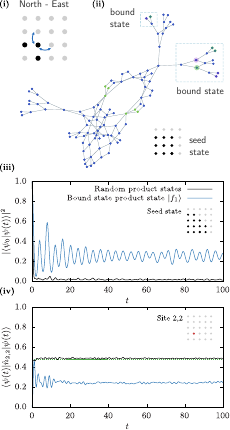}
    \caption{Bound states in the North-East model and their effect on dynamics. (i) Diagram of the allowed hoppings in the $U(1)$ North-East model. (ii) Graph of the largest sector with $N=9$ particles in $L=16$ sites with two ZM bound states, which are represented with the circle radii indicating the amplitude and the color (green or purple) indicating the phase ($\pi$ or $0$, respectively). Time evolution of the (iii) fidelity and the (iv) local occupation number  $\langle \hat{n}_{2,2}\rangle$ for $N=13$ particles in $L=25$ sites for the product state $|f_1\rangle$ belonging to the bound state in Eq.~\eqref{Eq2DBS} and the average of $50$ random product states. The dashed green line in (iv) represents the infinite temperature expectation value, around which the expectation value of the random states oscillates. The seed state that generates the sector and the measured site, $x=y=2$, are represented as insets.}
    \label{FigNorthEast}
\end{figure}

Here, we consider a 2D version of the $U(1)$ East model for a kinetic constraint range of $r=1$. A hard core boson can hop horizontally ($x$ axis) or vertically ($y$ axis) to its nearest neighbor if it has both a particle below and one to its left [see Fig.~\ref{FigNorthEast}(i)]. The Hamiltonian reads
\begin{equation}
    \hat{H}_{NE}=\sum_{x,y}\hat{n}_{x-1,y}\hat{n}_{x,y-1}(\hat{c}_{x,y}^{\dagger} \hat{c}_{x+1,y}+\hat{c}_{x,y}^{\dagger} \hat{c}_{x,y+1}+\text {h.c.}).\\
\end{equation}
This model also presents chiral symmetry, as it contains nearest neighbor hoppings that only couple opposite sublattices, which leads to a ZM subspace that can host bound states. Considering a square lattice, these sublattices form a checkerboard configuration, such that the operator 
\begin{equation}
\hat{\mathcal{C}}=\prod_{x,y=1}^{\lfloor L/2\rfloor}(2\hat{n}_{2x,2y}-1)\prod_{x,y=1}^{\lceil L/2\rceil}(2\hat{n}_{2x-1,2y-1}-1),
\end{equation}
anticommutes with the Hamiltonian. Similarly to its 1D counterpart, this model exhibits classical Hilbert space fragmentation, as any particle that does not have enough particles below or to its left is frozen. For example, all the lowest particles in each column and the leftmost particles in each row are frozen. 

In this model, we can consider the generalization of bound states to 2D. In the graph formalism, the concept is the same: a bound state is an eigenstate that is localized to a part of the graph which remains unchanged when increasing the size of the system, i.e., it has no weight on the set of vertices $V_2$, which acquire new edges in the larger graph. In the square lattice, the vertices in $V_2$ have a particle in the top row or the right column, which is mobile, with a particle below and one to its left, $\left| \substack{\bullet\ \bullet \\ *\ \bullet} \right\rangle$. Note that all states are robust to increasing the system to the left and to the bottom, as the particles in those boundaries are always frozen. Here, bound states can arise because the increase in the degrees of freedom in 2D is compensated by the stronger kinetic constraint, which reduces the number of edges in the graph and thus the number of vertices in $V_2$. 
Figure \ref{FigNorthEast}(ii) presents the graph of the largest sector of the system with $N=9$ particles in $L=16$ sites with colored vertices indicating the sets $V_1$ (blue) and $V_2$ (green). This sector is obtained from the domain wall seed state in 2D [see inset in \ref{FigNorthEast}(ii)].

The graph presents several tree structures that host zero modes compactly localized in a few vertices, similar to those appearing in Erdős-Rényi random graphs \cite{benami25}. As those states only populate the blue vertices, they are bound states. These tree structures naturally appear on the graph of this model due to its 2D nature: a mobile particle that can move to either the right or to the top, creates a bifurcation that can host a zero mode: 
$$\left| \substack{\circ\ \bullet\ \circ \\ \bullet\ \circ\ \circ \\ \circ\ \bullet\ \circ } \right\rangle \leftrightarrow
\left| \substack{\circ\ \circ\ \circ \\ \bullet\ \bullet\ \circ \\ \circ\ \bullet\ \circ } \right\rangle \leftrightarrow
\left| \substack{\circ\ \circ\ \circ \\ \bullet\ \circ\ \bullet \\ \circ\ \bullet\ \circ } \right\rangle 
\quad\Longrightarrow\quad \dfrac{\left| \substack{\circ\ \bullet\ \circ \\ \bullet\ \circ\ \circ \\ \circ\ \bullet\ \circ } \right\rangle-\left| \substack{\circ\ \circ\ \circ \\ \bullet\ \circ\ \bullet \\ \circ\ \bullet\ \circ } \right\rangle}{\sqrt{2}}.$$
Then, the combination of this configuration with other mobile particles generates more complex tree structures, as those depicted in Fig. \ref{FigNorthEast}(ii). For this case, $23$ out of $24$ ZMs are bound states due to the low number of vertices in $V_2$. 

Figures \ref{FigNorthEast}(iii,iv) illustrate the effect of these tree structures on a quench in a sector with $N=13$ particles in $L=25$ sites [see sector seed in Fig.~\ref{FigNorthEast}(iii)]. This sector hosts the following ZM bound state in a tree structure
\begin{equation}\label{Eq2DBS}
\dfrac{|f_1\rangle-|f_2\rangle}{\sqrt{2}}=\dfrac{\left| 
\substack{\circ\ \circ\ \bullet\ \circ\ \circ\  \\ \bullet\ \circ\ \circ\ \bullet\ \circ\  \\\bullet\ \circ\ \circ\ \bullet\ \circ\  \\\bullet\ \circ\ \bullet\ \bullet\ \bullet\ \\ \bullet\ \bullet\ \bullet\ \bullet\ \circ\  } \right\rangle-
\left| \substack{\circ\ \circ\ \bullet\ \circ\ \circ\  \\ \bullet\ \circ\ \circ\ \bullet\ \circ\  \\\bullet\ \bullet\ \circ\ \bullet\ \circ\  \\\bullet\ \circ\ \circ\ \bullet\ \bullet\ \\ \bullet\ \bullet\ \bullet\ \bullet\ \circ\ } \right\rangle}{\sqrt{2}}.
\end{equation}
We quench the system to the first of these two product states, $|f_1\rangle$, and compare their evolution to the averaged evolution of $50$ random product states in terms of the (iii) fidelity and the (iv) expectation value of a local density for a single site, $\langle \hat{n}_{2,2}\rangle$ (indicated in the inset). As $|f_1\rangle$ has a 50\% weight on the bound state, its fidelity remains around 0.25 throughout time evolution, while the fidelity of random product states rapidly decays to a small value close to zero as the states spread across the whole Hilbert space. Thus, a snapshot measurement of the whole system would show a high concentration on the particle configurations in Eq.~(\ref{Eq2DBS}), with measurement outcomes robust to increasing the size of the system. The expectation value $\langle \hat{n}_{2,2}\rangle$ for the random states reaches a saturation value around 0.48, which corresponds to the infinite temperature expectation value of the occupation of the site $x=y=2$ for that Hilbert space sector. In contrast, $\langle \hat{n}_{2,2}\rangle$ fluctuates around 0.25 for $|f_1\rangle$, as part of the population remains trapped in the bound state, which does not have this site occupied. Thus, the presence of bound states makes this model retain some memory of the initial state for some special initial states, which can be detected through local observables.

\subsubsection{Pair-flip model}
In this section, we discuss the pair-flip model, a 1D chain of hardcore bosons with inversion symmetry and without particle conservation. The Hamiltonian reads, 
\begin{equation}
    \hat{H}_{PF}=\sum_j \hat{n}_{j+1}\left(\hat{c}_j^{\dagger} \hat{c}_{j+2}^{\dagger}+\hat{c}_{j+2} \hat{c}_j\right),
\end{equation}
such that the only allowed process is $|{\circ\bullet\circ}\rangle \,\leftrightarrow \,|{\bullet\bullet\bullet}\rangle$: if there is a particle in the center site and the pair of adjacent sites have the same state, the state of this pair of sites can be flipped. This model is the dual of a modified version of the $U(1)$ East-West model, as it can be obtained by mapping the sites to the bonds of the original lattice \cite{aditya2024}. It also has chiral symmetry, $\{\hat{\mathcal{C}},\hat{H}_{PF}\}=0$, given by the operator $\hat{\mathcal{C}}=(-1)^{\hat{N}(\hat{N}-1)/2}$, where $\hat{N}=\sum_i\hat{n}_i$ is the total particle number operator. The strong kinetic constraint in this model yields a classical fragmentation pattern that can be characterized in terms of irreducible strings \cite{aditya2024}. As flips require the presence of a particle, the largest sector is generated by the seed configuration where all sites are occupied, as this maximizes the number of possible flips. 

As this model does not conserve the particle number, increasing the size of the system corresponds to adding either an empty or filled site. Table \ref{TabPairFlip} shows the configurations in the rightmost boundary that belong to the vertex sets $V_1$ and $V_2$, while those of the left boundary can be obtained by reflecting those configurations. The vertex always belongs to $V_1$ if the last site is empty, as this prevents dynamics involving the extra site. If the last site is occupied, the vertices in $V_1$ will be the ones where the added site is not in the same configuration as the second-to-last site, which also prevents dynamics.
\begin{table}[h]
    \centering
    \begin{tabular}{cc|cc}
    Add a filled site \hspace{2mm}& Vertex set\hspace{2mm} & Add an empty site & Vertex set\\\hline
        $|{\cdots\circ\bullet\rangle\otimes\color{gray}{|\,\bullet\rangle}}$ & $V_1$ & $|{\cdots\circ\bullet\rangle\otimes\color{gray}{|\,\circ\rangle}}$ &$V_2$ \\
        $|{\cdots\bullet\bullet\rangle\otimes\color{gray}{|\,\bullet\rangle}}$ & $V_2$ & $|{\cdots\bullet\bullet\rangle\otimes\color{gray}{|\,\circ\rangle}}$ & $V_1$ \\
        $|{\cdots*\circ\rangle\otimes\color{gray}{|\,\bullet\rangle}}$ & $V_1$ & $|{\cdots*\circ\rangle\otimes\color{gray}{|\,\circ\rangle}}$ & $V_1$ \\
    \end{tabular}
    \caption{Boundary conditions to generate bound states. Particle configurations in the vertex sets $V_1$ and $V_2$ for the right boundary of the pair-flip model.}
    \label{TabPairFlip}
\end{table}

\begin{figure}[t]
    \includegraphics[width=\columnwidth]{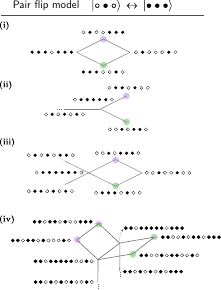}
    \caption{Simplest bound states in the pair-flip model. The circle radii indicate the amplitudes, and the colors (green or purple) indicate the phase ($\pi$ or $0$, respectively). The graph in (iv) hosts two states with non-zero energy, as represented by the upper and lower semicircles.}
    \label{FigPairFlip}
\end{figure}

The graph of the Hamiltonian exhibits many diamond and tree structures whose bifurcations reflect which pairs of sites are flipped. These structures can host localized eigenstates due to the presence of destructive interference, and, if only vertices in $V_1$ are populated, these states become bound states by sequentially adding the allowed configurations in Table \ref{TabPairFlip}. Figure \ref{FigPairFlip} depicts the smallest configurations of particles and their graphs that generate these localized states. If only two pairs of particles can be flipped, this generates a diamond structure [Fig.~\ref{FigPairFlip}(i)]. On the other hand, 5 consecutive filled sites generate a dangling tree structure [Fig.~\ref{FigPairFlip}(ii)], and 6 consecutive filled sites generate a dangling diamond [Fig.~\ref{FigPairFlip}(iii)]. One can combine these simple configurations with other unfrozen configurations to form more complex trees. These trees can host not only ZMs but also states at nonzero energy, which can also form bound states [see Fig.~\ref{FigPairFlip}(iv)]. 

\section{Discussion}\label{SecDiscussion}

In this work, we studied the zero mode (ZM) subspace in kinetically constrained models in the presence of both chiral symmetry and $U(1)$ conservation. First, we show that Hilbert space fragmentation notably increases the lower bound on the size of the degenerate subspace predicted by $U(1)$ and chiral symmetry. This yields a ZM subspace whose dimension increases exponentially with the size of the system. In order to study the structure of such a subspace, we propose a definition of a collective bound state which extends the notion of single-particle compact localized states~\cite{Flach2014a,Dias2015,Nandy2015,Pal2018,Ramachandran2017,MoralesInostroza2016,Mielke1991,Morfonios2021,Maimati2017,Maimati2021a,Maimati2021b,Maimati2021c,Xu2015,Sathe2021,Sathe2023,Rontgen2018} to many-body systems. We establish the sufficient conditions for the existence of these states and discuss how they might arise in degenerate subspaces. Finally, we also argue that bound states may be used as building blocks to construct factorizable eigenstates that feature zero entanglement across certain real-space cuts. 

We then demonstrate that our sufficient conditions for the existence of bound states are satisfied in two families of toy models, namely the $U(1)$ East and $U(1)$ East-West models. While both of the considered models feature $U(1)$ particle conservation, our construction is general and can be applied to other models. In order to demonstrate the generality of our construction, we briefly examine two additional models with quite distinct properties. In the North-East model, we show the existence of two-dimensional bound states, and in the pair-flip model, we show that bound states can also arise in the absence of a $U(1)$ symmetry related to particle conservation, and that they can have both zero and nonzero energy.  Furthermore, we demonstrate the influence of bound states and factorizable zero modes studied in our work on the physical properties of the system, such as dynamics from simple initial states. In this respect, they bear a parallel to quantum many-body scars~\cite{Turner2017,Serbyn2021Review,Chandran2023Review,Moudgalya2022Review}, but have a different physical origin. 

The models considered in our work are realizable on digital quantum platforms by mapping hardcore bosons to qubits and trotterizing the Hamiltonian. While time discretization may lead to additional overhead, we expect the dynamical signatures of simple bound states and Hilbert space quantum fragmentation to be robust with relatively large time steps. For example, the range $r=1$ $U(1)$ East model can be implemented with a brickwork circuit of unitary gates $U=e^{-i \theta\left[\hat{n}_j\left(c_{j+2}^{\dagger} c_{j+1}+\text {H.c.}\right)\right]}$, which correspond to a three-qubit control partial swap gate of the form $|0\rangle\!\langle 0|_c\otimes I_{ab} + |1\rangle\!\langle 1|_c\otimes e^{-i\frac{\theta}{2}(\sigma^x_a\sigma^x_b+\sigma^y_a\sigma^y_b)}$, where $\sigma^{x,y}_{a,b}$ denote Pauli matrices, also known as a generalized Fredkin gate \cite{patel2016}. This gate is realizable on digital quantum platforms, either natively, in neutral (Rydberg) atom arrays \cite{khazali2025}, or by decomposition, using superconducting qubits \cite{krizan2025}.

\begin{table*}[thb!]
    \begin{center}
        \begin{tabular}{c|c c c c |c c c c|c c c c}
            \hline\hline\rule{0pt}{12pt}
             & $r=1$  & & &  & $r=2$ &   & & & $r=3$ &  &  & \\[1mm]\hline
            $N$ & $L$ & $\mathcal{M}$ & $N_{ZM}$ & $N_{FS}$ & $L$ & $\mathcal{M}$ & $N_{ZM}$ & $N_{FS}$ & $L$ & $\mathcal{M}$ & $N_{ZM}$ & $N_{FS}$\\[1mm]\hline
            2  & 3  & 0  & 0  & 0 & 4  & 1  & 1   & 0 & 5  & 0 & 0  & 0\\
            3  & 5  & 1  & 1  & 0 & 7  & 2  & 2   & 0 & 9  & 2 & 2  & 0\\
            4  & 7  & 0  & 0  & 0 & 10 & 3  & 7   & 2 & 13 & 0 & 4  & 2\\
            5  & 9  & 2  & 2  & 0 & 13 & 7  & 11  & 6 & 17 & 9 & 11 & 3\\
            6  & 11 & 0  & 4  & 1 & 16 & 12 & 22  & 15 & 21 & 0 & 12 & 8\\
            7  & 13 & 5  & 7  & 3 & 19 & 30 & 58  & 39 & 25 & 52 & 68 & 20\\
            8  & 15 & 0  & 4  & 2 & 22 & 55 & 127 & 91 & 29 & 0 & 62 & 56\\
            9  & 17 & 14 & 16 & 3 &  25&  143&  315&  210&  &  &  &\\
            10 & 19 & 0  & 10 & 6 &  &  &  & &  &  &  &
        \end{tabular}
    \end{center}
    \caption{Number of ZMs, $N_{ZM}$, and factorizable eigenstates, $N_{FS}$, in the largest sector of the $U(1)$ East model with ranges $r=1,2,3$. For each number of particles we consider $L=(r+1)N-r$ sites and we indicate the mismatch after fragmentation in the considered sector, $\mathcal{M}$. }
    \label{TabEast}
\end{table*}

\begin{table*}[thb!]
    \begin{center}
        \begin{tabular}{c|c c c c|c c c c|c c c c}
            \hline\hline\rule{0pt}{12pt}
             & $N=4$ &  &  & & $N=5$ & & & & $N=6$ & \\[1mm]\hline
            $L$ & $\mathcal{M}$ & $N_{ZM}$ & $N_{BS}$ & $N_{FS}$ & $\mathcal{M}$ & $N_{ZM}$ & $N_{BS}$ & $N_{FS}$ & $\mathcal{M}$ & $N_{ZM}$ & $N_{BS}$ & $N_{FS}$\\[1mm]\hline
            10 & 10 & 10 & 1 & 0 & 0 & 0 & 0 & 0 & 10 & 10 & 0 & 0 \\
            11 & 10 & 14 & 5 & 3 & 10 & 10 & 0 & 0 & 10 & 10 & 0 & 0 \\ 
            12 & 15 & 27 & 15 & 12 & 0 & 0 & 0 & 0 & 20 & 20 & 0 & 0 \\ 
            13 & 14 & 46 & 34 & 30 & 15 & 15 & 1 & 0 & 20 & 20 & 0 & 0 \\ 
            14 & 20 & 80 & 65 & 61 & 0 & 6 & 6 & 4 & 35 & 35 & 0 & 0 \\
            15 & 18 & 126 & 111 & 107 & 21 & 39 & 21 & 18 & 35 & 35 & 0 & 0 \\ 
            16 & 25 & 193 & 175 & 171 & 0 & 56 & 56 & 52 & 56 & 56 & 1 & 0 \\ 
            17 & 22 & 278 & 260 & 256 & 27 & 147 & 125 & 121 & 56 & 62 & 7 & 5 
        \end{tabular} 
    \end{center}
    \caption{Number of ZMs, $N_{ZM}$, bound states, $N_{BS}$, and factorizable eigenstates, $N_{FS}$, in the largest sector of the $U(1)$ East-West model for $N=4,5,6$ particles and $r=2$. For each number of particles we consider $L=[10-17]$ sites and we indicate the mismatch after fragmentation in the considered sector, $\mathcal{M}$.}
    \label{TabWest}
\end{table*}

Bound states also naturally connect to Hilbert space fragmentation~\cite{Moudgalya2022Review} and may provide new insights into this phenomenon. For example, in the $U(1)$ East model, it is possible to differentiate between classical and quantum fragmentation by considering the entire family of bound states on some fixed number of sites. If the space spanned by these bound states admits a product state basis, then the construction leads to classical fragmentation; if it does not, then it leads to quantum fragmentation. Furthermore, the same mechanism works in the PXP model and charge and dipole conserving models, where bound states are simply the eigenstates for which adding an extra site results in an inert configuration,
thus leading to fragmentation. Therefore, studying bound states in other models could lead to a new approach to both finding and distinguishing between classical and quantum fragmentation.  

Additionally, dynamical symmetries, corresponding to operators $\hat J$, that fulfill the following commutation relation with the Hamiltonian, $[\hat{H},\hat{J}]=\omega\hat{J}$ \cite{b1988dynamical,buca2019a,gunawardana2022,moudgalya2020}, naturally arise from the collective bound states and factorizable states. Take the projector $\hat{J}=|\psi_1^{\ell}\rangle\langle\psi_2^{\ell}|\otimes|{\circ\circ\circ}\rangle\langle\circ\circ\circ|\otimes \mathbb{1}$, built from two bound states with the same number of particles and sites, a padding configuration, and the identity acting on the remaining sites on the right. This projector, which is localized on a few sites, commutes with the Hamiltonian, $[\hat{H},\hat{J}]=0$, and is thus a localized dynamical symmetry with $\omega=0$. Dynamical symmetries have been shown to lead to persistent oscillations and can be used to compute autocorrelation functions \cite{buca2025}, out-of-time-ordered correlators \cite{medenjak2020,buca2022a}, and quenches \cite{buca2023}. It is an interesting open question if the dynamical symmetries emergent from collective bound states can be used to obtain new analytical insights into the dynamics of the $U(1)$ East models.

\emph{Note added.} During the final stage of preparing this manuscript, two related works~\cite{tan25,benami25} appeared on arXiv, while a third work appeared shortly afterwards~\cite{jonay2025}. These studies investigate analogues of compact localized states in many-body quantum systems. Using the terminology introduced in our work, these states may be potential candidates for realizing collective bound states, provided they remain stable under spatial extension of the system.

\begin{acknowledgments}
The authors acknowledge useful discussions with Berislav Buča. This work was supported by the European Research Council (ERC) under the European Union's Horizon 2020 research and innovation program (Grant Agreement No.~850899).
M.~L. acknowledges support by the Deutsche Forschungsgemeinschaft (DFG, German Research Foundation) under Germany’s Excellence Strategy – EXC-2111 – 390814868. 
This research was supported in part by grant NSF PHY-2309135 to the Kavli Institute for Theoretical Physics (KITP). 
\end{acknowledgments}

\appendix

\section{Counting of zero modes}\label{SecAppendix}
In the main text, we have considered particular cases of the $U(1)$ East and $U(1)$ East-West models. Here, we illustrate how the appearance of bound states and factorizable eigenstates in the ZM subspace generalizes to other instances of these families of models. Table \ref{TabEast} gives the number of  factorizable ZMs for the $U(1)$ East model with ranges $r=1,2,3$. We consider the largest sector of the Hilbert space for different numbers of particles, $N$, and numbers of sites $L=(r+1)N-r$. We indicate the mismatch $\mathcal{M}$ within the largest sector, the number of ZMs, $N_{ZM}$, and the number of factorizable eigenstates $N_{FS}$. For this particle filling, all ZMs are bound states, and all ranges, $r=1,2,3$, exhibit ZM factorizable eigenstates, with their numbers increasing with the number of particles for both $r=2,3$. Table~\ref{TabWest} presents the ZM state counts for the largest sector of the $U(1)$ East-West model with $r=2$, i.e., excluding the frozen states. As this model does not have a natural lengthscale, we consider a range of system sizes $L\in[10,17]$ for each number of particles, $N=4,5,6$. For both $N=4$ and $N=5$, the number of bound states and factorizable eigenstates increases with the size of the system. In both families of models, bound states and factorizable eigenstates can exist even for $\mathcal{M}=0$, as a nonzero mismatch only provides a lower bound to the number of ZMs.

\begin{figure}[t]
    \includegraphics[width=\columnwidth]{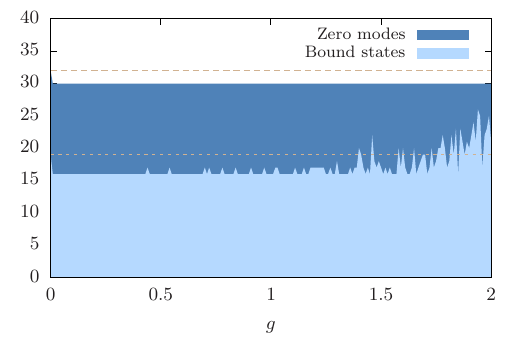}
    \caption{Robustness of bound states and zero modes against tunneling disorder for the $U(1)$ East model. We depict the number of bound states and zero modes for a single random disorder realization with $r=2$, $N=7$ particles in $L=18$ sites as a function of the disorder strength $g$. The dashed horizontal lines indicate the number of unperturbed zero modes and bound states.}
    \label{FigDisorder}
\end{figure}

\section{Bound state and zero mode robustness}\label{SecAppendix2}

In this Appendix, we examine the stability of zero modes and bound states in the $U(1)$ East model against perturbations. We consider two classes of perturbations that preserve chiral symmetry: perturbations that preserve the kinetic constraints and perturbations that do not. As an example of a perturbation that does not affect the kinetic constraints, one can introduce randomness in the hopping terms. The Hamiltonian for the range 2 model with random tunneling terms reads
\begin{equation}
\hat{\tilde{H}}_{2,E}=\sum_{i=2}^{L-1}(1+g\omega_i)(\hat{n}_{i-1}+\hat{n}_{i-2}-\hat{n}_{i-1} \hat{n}_{i-2})(\hat{c}_i^{\dagger} \hat{c}_{i+1}+\text {h.c.}),
\end{equation}
where $\omega_i\in[-0.5,0.5]$ is a random number that we draw from a uniform distribution, and $g$ is the disorder strength. Figure \ref{FigDisorder} shows the number of zero modes and bound states as a function of the perturbation strength $g$ for the range $r=2$ model with $N=7$ particles in $L=18$ sites. As an example, we show a typical disorder realization, which shows how both types of states are robust to tunneling disorder. The number of zero modes only decreases from 32 to 30 and stays constant at this value for any nonzero disorder strength. Similarly, the number of bound states decreases from 19 to 16, but it exhibits fluctuations above the lower threshold, which increase for strong disorder strengths away from the perturbation regime. 

\begin{figure}[t]
    \includegraphics[width=\columnwidth]{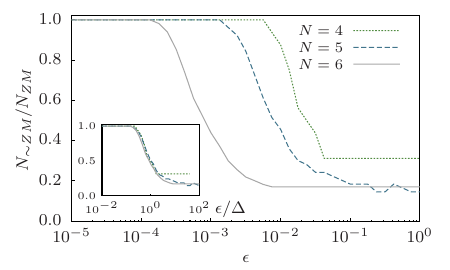}
    \caption{Lifting of the degeneracy of the ZM subspace for the $U(1)$ East model with $r=2$ with an uncorrelated hopping perturbation. Number of states with energy $E\in[-\Delta,+\Delta]$, $N_{\sim ZM}$, where $\Delta$ is the average level spacing of the excited states close to zero energy in the unperturbed model normalized to the number of unperturbed ZMs, $N_{ZM}$, as a function of the perturbation strength $\epsilon$. The inset shows the collapse of the curves with different numbers of particles $N$ for a perturbation strength rescaled by the average level spacing $\Delta$.}
    \label{FigLifting}
\end{figure}

As an example of a perturbation that affects the kinetic constraints, we add a term to the Hamiltonian that implements uncorrelated hopping. The range $r=2$ case reads,
\begin{equation}
\begin{aligned}
    \hat{\tilde{H}}_{2,E}=&\sum_{i=2}^{L-1}\Big[(\hat{n}_{i-1}+\hat{n}_{i-2}-\hat{n}_{i-1} \hat{n}_{i-2})(\hat{c}_i^{\dagger} \hat{c}_{i+1}
    +\text {h.c.})+\\&\epsilon(\hat{c}_{i+1}^{\dagger} \hat{c}_i+\text {h.c.})\Big],
\end{aligned}
\end{equation}
where $\epsilon$ is the strength of the perturbation. For any $\epsilon>0$, the Hilbert space is no longer fragmented, which partially lifts the degeneracy of the ZM subspace. In Fig. \ref{FigLifting}, we inspect this process by considering an energy window, $[-\Delta,+\Delta]$, where $\Delta$ is the mean-level spacing for highly excited states close to zero energy in the unperturbed case. We plot the number of states within this energy window normalized to the number of unperturbed ZMs, $N_{\sim ZM}/N_{ZM}$, as a function of the perturbation strength $\epsilon$ for different numbers of particles. The inset shows the collapse of the curves when the perturbation strength is scaled to the mean-level spacing, which decreases exponentially with the system size. As the perturbation preserves chiral symmetry, the number of states $N_{\sim ZM}$ does not decay to one, as it is lower bounded by the chiral mismatch [Eq.~(\ref{EqMismatch})]. In all cases, this lower bound of ZMs is saturated, which we also expect to hold for larger system sizes. As the mismatch grows exponentially with the system size but at a lower rate than the number of ZMs in the unperturbed $U(1)$ East model (see Fig.~\ref{FigMismatch}), we expect the ratio $N_{\sim ZM}/N_{ZM}$ to go to zero for $N\rightarrow\infty$. 

In summary, hopping disorder preserves both chiral symmetry and kinetic constraints and barely affects the number of ZMs and bound states of the model. Uncorrelated hopping only preserves chiral symmetry but not the kinetic constraints, such that the enhancement of the number of ZMs given by the fragmentation in the unperturbed $U(1)$ East model is lost. However, as discussed in the main text, the initial states with a high overlap with a bound state and factorizable eigenstate exhibit fidelity revivals even in the presence of uncorrelated hopping. For more generic perturbations that might not preserve the chiral symmetry, bound states are not expected to be robust. For example, a random on-site potential of the form $g\omega_i\sum_i \hat{n}_i$ removes chiral symmetry, which makes the zero-mode subspace that hosts the bound states disappear. Still, the dynamical signatures of bound states might be resilient to small enough chirality-breaking perturbations as they are for the case of uncorrelated disorder.

\bibliography{references}

\end{document}